\documentclass[lettersize,journal]{IEEEtran}
\usepackage{amsmath,amsfonts}
\usepackage{algorithmic}
\usepackage{algorithm}
\usepackage{array}
\usepackage{tabulary}
\usepackage[caption=false,font=normalsize,labelfont=sf,textfont=sf]{subfig}
\usepackage{textcomp}
\usepackage{stfloats}
\usepackage{url}
\usepackage{verbatim}
\usepackage{graphicx}
\usepackage{cite}
\usepackage{xcolor}

\begin{document}

\title{Enhancing Decision-Making in Windows PE Malware Classification During Dataset Shifts with Uncertainty Estimation}

\author{Rahul Yumlembam, Biju Issac, and Seibu Mary Jacob
        % <-this % stops a space
\thanks{Rahul Yumlembam, and Biju Issac are with the Department of Computer and Information Sciences, Northumbria University, Newcastle, UK (email: r.yumlembam@northumbria.ac.uk, bijuissac@northumbria.ac.uk). Corresponding author: Biju Issac}% <-this % stops a space
\thanks{Seibu Mary Jacob is with the School of Computing, Engineering \& Digital Technologies, Teesside University, Middlesbrough, UK (email: s.jacob@tees.ac.uk).}
}

% The paper headers
\markboth{Journal of \LaTeX\ Class Files,~Vol.~14, No.~8, August~2021}%
{Shell \MakeLowercase{\textit{et al.}}: A Sample Article Using IEEEtran.cls for IEEE Journals}

% \IEEEpubid{0000--0000/00\$00.00~\copyright~2021 IEEE}
% Remember, if you use this you must call \IEEEpubidadjcol in the second
% column for its text to clear the IEEEpubid mark.

\maketitle

\begin{abstract}
Artificial intelligence techniques have achieved strong performance in classifying Windows Portable Executable (PE) malware, but their reliability often degrades under dataset shifts, leading to misclassifications with severe security consequences. To address this, we enhance an existing LightGBM (LGBM) malware detector by integrating Neural Networks (NN), PriorNet, and Neural Network Ensembles, evaluated across three benchmark datasets: EMBER, BODMAS, and UCSB. The UCSB dataset, composed mainly of packed malware, introduces a substantial distributional shift relative to EMBER and BODMAS, making it a challenging testbed for robustness. We study uncertainty-aware decision strategies, including probability thresholding, PriorNet, ensemble-derived estimates, and Inductive Conformal Evaluation (ICE).
Our main contribution is the use of ensemble-based uncertainty estimates as Non-Conformity Measures within ICE, combined with a novel threshold optimisation method. On the UCSB dataset, where the shift is most severe, the state-of-the-art probability-based ICE (SOTA) yields an incorrect acceptance rate (IA\%) of 22.8\%. In contrast, our method reduces this to 16\% a relative reduction of about 30\% while maintaining competitive correct acceptance rates (CA\%). These results demonstrate that integrating ensemble-based uncertainty with conformal prediction provides a more reliable safeguard against misclassifications under extreme dataset shifts, particularly in the presence of packed malware, thereby offering practical benefits for real-world security operations.
\end{abstract}

\begin{IEEEkeywords}
Conformal Prediction, Windows PE Malware, Machine Learning, Deep Learning, Uncertainty estimation
\end{IEEEkeywords}

\section{Introduction} \label{Introduction}
Artificial Intelligence (AI) and machine learning are pivotal in enhancing the detection, prevention, and response mechanisms against Windows Portable Executable (PE) malware in cybersecurity. The Portable Executable (PE) format serves as the standard file format for executable programs within Microsoft's 32-bit and 64-bit Windows operating systems \cite {Sikorski}. PE files encompass various formats, including .exe files, dynamic link libraries (.dlls), BAT/Batch files (.bat), control panel applications (.cpl), kernel modules (.srv), device drivers (.sys), and numerous others, with .exe files being the most prevalent among them. Recent studies illustrate that machine learning and deep learning models, characterized by various representations of PE files, have achieved high accuracy in malware detection. However, high accuracy only sometimes translates to reliability, especially in new or ambiguous cases. Typically, classifiers predict the class of an instance based on the highest probability, neglecting the confidence or uncertainty of this prediction. For instance, if a model predicts a PE file as benign with a 60\% probability and as malware with a 40\% probability, it classifies the file as benign despite the narrow margin, indicating low confidence in the prediction. Such borderline predictions pose risks in critical applications like cybersecurity, prompting the need for strategies to assess each prediction's certainty. For example, a benign PE file such as a Windows update utility may be misclassified as benign with low confidence, while actually embedding a trojan downloader. This misclassification allows it to bypass security layers and trigger further payload downloads, leading to severe security breaches in enterprise networks. There are two broad categories which induce uncertainty in the classifier's prediction: (1) Data uncertainty (Aleatoric Uncertainty), caused by measurement errors or inherent system randomness, and (2) Model uncertainty (Epistemic Uncertainty), resulting from inadequate knowledge or model limitations~\cite{gawlikowski2023survey}. Addressing model uncertainty is feasible through improvements in architecture, learning processes, or training data, while data uncertainty is irreducible ~\cite{kendall2017}.

Consider a scenario where a security operations centre deploys a machine learning model trained on historical malware samples. An attacker develops a slightly modified ransomware variant that incorporates new obfuscation techniques not seen during training. The model may classify the sample as benign with 55\% probability and as malware with 45\% probability, reflecting low confidence. In such borderline cases, relying solely on the highest-probability prediction may lead to operational risk, either through missed detections or false alarms, both of which can have serious financial or security consequences. Uncertainty estimation helps identify such cases and allows for appropriate human intervention or automated safeguards. In practice, this means that low-confidence predictions are not treated as final outcomes, but instead routed through different escalation paths. For example, in Security Operations Centers (SOCs), uncertain classifications can be flagged for manual review by analysts, who may conduct reverse engineering, inspect suspicious API calls, or correlate the file with external threat intelligence feeds. Analysts can also prioritize these flagged files for deeper forensic investigation, reducing the chance of a novel malware sample bypassing detection. On the automated side, high-uncertainty predictions can trigger additional protective measures. These may include detonation of the file in a sandbox or virtualized environment to observe dynamic behavior, application of heuristic or signature-based scans from complementary tools, or temporary isolation of the executable by quarantining it until further verification is performed. In enterprise deployments, such mechanisms may also activate layered defenses, such as Endpoint Detection and Response (EDR) policies that restrict file execution, or automated network segmentation to contain potential compromise. By incorporating these escalation strategies, uncertainty estimation provides a practical mechanism to balance automation with human expertise, reducing operational risks and enhancing resilience against novel or obfuscated threats.

For example, a file may be classified as benign with only a narrow margin of confidence (e.g., 55\% benign vs 45\% malware), which can pose substantial risk if misclassified. In such borderline cases, quantifying predictive uncertainty allows security systems to either defer decision-making or trigger additional analysis, improving robustness against evolving threats. This work focuses on Windows PE malware since it represents the majority of malware infections in enterprise and consumer endpoints worldwide. PE files cover a wide spectrum of malware types, including ransomware, trojans, and spyware, making them ideal for uncertainty estimation studies. Android and Linux malware may exhibit different structural characteristics requiring dedicated models and datasets, which are reserved for future work.

Previous research has explored uncertainty estimation in malware detection using various techniques, but recent advancements such as PriorNet \cite{malinin2019reverse}, along with Neural Network Ensemble and Conformal Prediction for uncertainity estimation \cite{jordaney2017transcend}\cite{barbero2022transcending } remain less examined. Moreover, studies often overlook the importance of evaluating uncertainty estimation during dataset shifts — a critical oversight, as estimates can be misleading if not tested under these conditions. Furthermore, existing literature focusing primarily on traditional performance metrics may not fully capture the implications of rejecting predictions, especially the rejection of correct predictions.

In our paper, we enhance an existing LightGBM (LGBM) based malware detector ~\cite{anderson2018EMBER} \cite{yang2021bodmas}, integrating it with Neural Network (NN), PriorNet~\cite{malinin2019reverse}, and an Ensemble of Neural Networks to improve F1 scores. We then calibrate the probabilities of the LGBM model using Isotonic Regression and the Neural Network model using Temperature Scaling. Predictions are then accepted or rejected based on classifier probabilities threshold, uncertainty estimates threshold (Expected Entropy, Entropy of Expected, and Knowledge Uncertainty) from PriorNet and Neural Network Ensemble, and P-values from Inductive Conformal Evaluation. We make the following novel contributions through this research.

\begin{itemize}
    \item We proposed to utilize the uncertainty estimate from NN-Ensemble, such as Expected Entropy, Entropy of Expected, and Knowledge Uncertainty, into the Inductive Conformal Evaluation (ICE) framework to improve the rejection of incorrect prediction during dataset shift, while using three PE malware datasets, i.e., EMBER, UCSB and BODMAS.
    \item We proposed a unique threshold optimization that effectively balances acceptance of correct prediction and rejection of incorrect prediction.
    \item This is the first work to extensively evaluate PriorNet in the context of PE Malware detection.
    \item In previous studies, the state-of-the-art method \cite{barbero2022transcending} was evaluated using the EMBER 2018 dataset, with the first five months designated for training and the subsequent months for testing. This approach might not provide a comprehensive understanding of model performance, particularly in terms of simulating dataset shifts. In this work, we have adopted a more rigorous approach by not only testing on the same split from the EMBER dataset but also extending our evaluation to include the UCSB dataset, which contains packed malware, and the BODMAS 2020 dataset. This multi-dataset testing strategy allows us to assess how well the model trained on EMBER adapts to new and diverse datasets, offering a more detailed picture of its robustness against dataset shifts.
    \item In contrast to prior work, such as Barbero et al. [6], which uses probability scores as NCM, our work introduces uncertainty estimates derived from neural network ensembles as alternative NCMs. This approach better captures epistemic uncertainty under dataset shifts. Furthermore, we propose a novel threshold optimisation strategy using harmonic mean to balance the trade-off between correctly accepted and correctly rejected predictions, ensuring both high confidence and safety in operational environments.
\end{itemize}

We comprehensively analyzed the acceptance and rejection of correct and incorrect predictions using datasets with and without dataset shifts. Our experiments yield several insights:
\begin{itemize}
    \item Calibrating LGBM with Isotonic Regression and Neural Network with Temperature Scaling enhances overall classification performance and reduces incorrect prediction acceptance across various techniques.
    %\item  Without dataset shifts, state-of-the-art technique ICE with probability as NCM reduces incorrect prediction acceptance across various techniques.
    \item  During dataset shifts, the proposed method of using uncertainty estimates, particularly Expected Entropy and Entropy of Expected from Neural Network Ensemble as NCM, significantly reduce incorrect prediction acceptance, outperforming state-of-the-art technique \cite{barbero2022transcending} that uses ICE with probability as NCM.
    \item Across the datasets, there was an improvement in the rejection of Incorrect Prediction.
    \item The Ensemble Method is better in performance than PriorNet during dataset shifts. We suspect that the need for out-of-distribution (OOD) data is the primary reason, as PriorNet heavily relies on it to generate uncertain estimates for OOD data. %Unfortunately, we cannot predict what OOD data will entail.
\end{itemize}

Unlike prior works that either apply probability-based thresholds or rely solely on ICE with basic probability non-conformity measures, our method integrates neural network ensemble uncertainty into ICE, achieving finer discrimination between correct and incorrect predictions, especially under dataset shift. Our novel harmonic threshold optimisation ensures a better balance between correctly accepted and correctly rejected predictions compared to fixed rejection rate methods.

The remainder of this paper is organized as follows. Section II reviews related works on malware detection and uncertainty estimation. Section III states the problem setup and research questions. Section IV details our empirical analysis methodology, including datasets, feature extraction, classifier choices, uncertainty estimation techniques, and their integration with Inductive Conformal Evaluation (ICE). Section V presents experimental results and analysis and it answers the research questions (RQ1–RQ4) based on the empirical findings. Finally, Section VI concludes and outlines future directions.

\section{Related Works} \label{related_works}

Recent works have increasingly focused on leveraging uncertainty estimation to improve malware detection performance, especially under dataset shift. Papadopoulos et al.\ \cite{Papadopoulos2023} proposed an Inductive Conformal Prediction approach with unbiased confidence guarantees for Android malware detection, demonstrating the value of conformal methods in security applications. Li et al.\ \cite{Li2021} studied predictive uncertainty in Android malware classifiers and showed how uncertainty estimation can assist in detecting both dataset shift and adversarial examples. To address general uncertainty estimation under shift, Zaidi et al.\ \cite{Zaidi2021} introduced Neural Ensemble Search, which automatically optimizes ensemble architectures for improved uncertainty quantification. Rivera et al.\ \cite{Rivera2024} proposed an efficient conformal prediction framework for ensembles, employing score-based aggregation to improve calibration and predictive confidence.  In a related direction, Doan et al.\ \cite{Doan2024} demonstrated that Bayesian models can detect adversarial malware without requiring explicit adversarial training, highlighting the potential of Bayesian uncertainty for robust malware detection. These recent advances strongly motivate our investigation into combining ensemble-based uncertainty estimation with conformal prediction for Windows PE malware classification under dataset shifts.

Many approaches using machine learning and deep learning have been proposed to detect PE malware, demonstrating significant advancements. Anderson et al.\cite{anderson2018EMBER} made strides by extracting static features from PE files, such as headers and imported and exported functions, and employing a gradient-boosted decision tree with LGBM as the classifier. Building on this, Shiva et al.\cite{shiva2019windows} utilized static and dynamic information to classify malware, opting for a Linear Support Vector Machine as the classifier. Sun et al.\cite{sun2019opcode} introduced a novel method by analyzing opcode frequency and employing the K-nearest neighbour algorithm to categorize PE files into malware or benign based on similarity.
The role of Application Programming Interface (API) calls, and their sequences have been a focal point in understanding software behaviour, with several studies highlighting their significance. For instance, S. L. et al.\cite{sl2019windows} innovated by generating images from sequences of n bytes derived from API calls, utilizing a Convolutional Neural Network (CNN) for classification. Similarly, Amer et al.\cite{amer2020dynamic} explored the contextual traits of API call sequences to identify PE malware effectively.
Furthermore, the research conducted by Rabadi et al.\cite{rabadi2020advanced} took into consideration API calls and their arguments, indicating a more detailed approach. Parildi et al.\cite{parildi2021deep} proposed the use of Word2Vec for opcode embedding generation, training various deep learning models such as CNN, Transformer, LSTM, and traditional machine learning algorithms like Logistic Regression and Support Vector Machine, to enhance detection capabilities. Xu et al.\cite{xu2021malbert} suggested pretraining with BERT and fine-tuning using labelled API sequence data. In contrast, Ling et al.\cite{ling2022malgraph} employed a graph to represent relational information among functions, using a Graph Neural Network (GNN) for classification, illustrating the diversity in methodological approaches to PE malware detection. The broader field of intrusion detection has also explored the integration of machine learning methods to enhance system-level security. For example, Elzaridi and Kurnaz \cite{elzaridi2024integration} discuss how coupling intrusion detection with machine learning techniques can optimize network security. While their focus is on IDS integration at the system level, our study is concerned with classifier-level uncertainty estimation and rejection mechanisms for Windows PE malware detection. As classifiers are prone to errors, researchers have been looking for a way to reject incorrect predictions, and the most common approach is to use Uncertainty Estimation.
Regarding uncertainty estimation for malware classifier, Li et al.\cite{li2021can} applied Monte Carlo dropout, Variational Bayesian Inference, and Deep Ensemble on an Android Malware dataset. However, their analysis needed a comprehensive evaluation of correctly accepted and rejected predictions, leaving room for further exploration. Using an Ensemble approach, Nguyen et al.\cite{nguyen2021leveraging} assessed uncertainty estimation for PE malware. However, they did not compare it against other methods like probability thresholding or post-processing techniques such as conformal prediction. Their analysis focused on the True Positive Rate (TPR) and False Positive Rate (FPR) post-rejection, which only partially conveys the efficacy of prediction acceptance or rejection, especially under conditions of dataset shift. Barbero et al.\cite{barbero2022transcending} introduced an approximate Transductive and Inductive Conformal Evaluator based on Conformal Prediction, with a primary focus on the Android dataset Androzoo\cite{allix2016androzoo}. While their research highlights the potential of Conformal Prediction in malware classification, testing the EMBER v2 dataset across different timelines may only partially simulate fundamental dataset shifts. A more comprehensive analysis across multiple datasets could reveal more profound insights, particularly with known dataset shifts. Furthermore, using the probability outputs from a single model as the Non-Conformity Measure (NCM) can often be misleadingly overconfident, which is especially true in scenarios where the training data do not fully represent the feature space or contain noisy labels, suggesting opportunities for improving the NCM. Beyond these focused methodological contributions, Elayni et al.\ \cite{moamin2025artificial} provide a broad survey of artificial intelligence techniques, datasets, and challenges in malware and network intrusion detection. While their survey does not specifically address uncertainty estimation or conformal prediction, it situates our work within the broader AI-driven cybersecurity literature. Similarly, Yaseen et al.\ \cite{yaseen2025enhanced} explored metaheuristic-based feature selection and SVM optimization for intrusion detection, highlighting complementary advances in classifier optimization.

Although these studies provide valuable advances in malware detection, they leave several important gaps unaddressed. Prior works have either relied on probability-based thresholds or employed conformal prediction with basic non-conformity measures, often overlooking richer ensemble-derived uncertainty metrics. Moreover, few studies have systematically evaluated the rejection of incorrect predictions under explicit dataset shifts, which is critical in realistic malware scenarios where adversaries continuously evolve. Approaches such as PriorNet and single-model uncertainty estimation offer lightweight solutions but struggle to generalize without carefully crafted out-of-distribution data. Likewise, existing surveys and IDS-focused research highlight broader methodological trends but do not explore classifier-level uncertainty calibration or rejection strategies tailored to PE malware. These limitations directly motivate our approach, which seeks to go beyond these constraints.

For instance, Barbero et al. \cite{barbero2022transcending} introduced Inductive Conformal Evaluation (ICE) using probability-based Non-Conformity Measures (NCM) and focused on Android malware (AndroZoo). Our work significantly extends this approach by integrating richer uncertainty metrics derived from Neural Network Ensembles—namely, Expected Entropy, Entropy of Expected, and Knowledge Uncertainty—into the ICE framework. This allows for a more nuanced and robust characterisation of epistemic uncertainty, particularly under dataset shift conditions. Moreover, unlike Barbero et al. [6] who limited their evaluation primarily to Android datasets, we conduct extensive experiments on multiple independent Windows PE malware datasets (EMBER, UCSB, BODMAS), including packed malware scenarios, to provide more comprehensive and practically relevant insights for Windows security ecosystems. Furthermore, unlike prior studies [6][19][20], which relied on either single-model probability outputs or basic ensemble metrics, we uniquely combine ensemble-based uncertainty with ICE while also introducing a novel harmonic threshold optimization strategy that better balances the trade-off between correctly accepted and correctly rejected predictions an aspect not addressed in previous ICE implementations.

\section{Problem Statement} \label{problem_statement}
%Machine learning models predict new data instances, \(x'\), by applying the generalization principles learned from training data, \(x_i\). This generalization inherently introduces Uncertainty, attributable to factors such as noise during data acquisition, the process of building the machine learning model, and the choice of inference methods, like ensembles of neural networks. These Uncertainties are broadly classified into Aleatoric and Epistemic. Aleatoric Uncertainty, rooted in the data acquisition process, is due to randomness or inherent variability in the modelled system and is immutable despite additional data. In contrast, Epistemic Uncertainty pertains to the model's data comprehension. It can be diminished through further data collection, model refinement, or the adoption of more sophisticated inference techniques, given that it stems from the model's limitations in comprehensively learning the underlying data distribution. 

This paper delves into Epistemic Uncertainty, particularly in malware detection—a field where the adversary continuously evolves, rendering static feature extraction and prediction methods inadequate ~\cite{gawlikowski2023survey}.

In today's ever-evolving threat landscape, malware is becoming increasingly complex and challenging to predict. However, by utilizing uncertainty estimation, we can enhance the reliability of our predictions. This study delves into various strategies for achieving more accurate and dependable malware predictions by estimating uncertainty. Traditional approaches have not fully explored the effects of applying thresholds to predictions based on uncertainty estimates—specifically, their impact on the acceptance or rejection of predictions. Moreover, previous works have not extensively studied newer techniques like PriorNet and Inductive Conformal prediction. Therefore, to answer some of these questions, our empirical study is centred around answering the following research questions:
\textbf{RQ1}: Does calibrating the logits from classifiers reduce the acceptance of incorrect predictions, thereby enhancing prediction reliability?
\textbf{RQ2}: In the absence of dataset shift, what is the efficacy of using probabilities, uncertainty estimates, and  Inductive Conformal Evaluation in minimizing the acceptance of incorrect predictions?
\textbf{RQ3}: When faced with dataset shifts, can these methods maintain the accuracy of prediction by reducing incorrectly predicted instances?
\textbf{RQ4}: Can a single deterministic uncertainty estimation technique, such as PriorNet and Inductive Conformal Evaluation, provide a cost-effective yet reliable alternative to resource-intensive ensemble methods?

The hypotheses underlying these research questions are as follows:
H1 (RQ1): Calibration of classifier outputs reduces the acceptance of incorrect predictions without severely sacrificing correct predictions.

H2 (RQ2): Uncertainty-based thresholds (using uncertainty metrics or ICE) outperform simple probability thresholds in minimizing incorrect predictions in stationary (no-shift) data.

H3 (RQ3): Under dataset shift, uncertainty estimates derived from neural ensemble models retain their ability to distinguish correct from incorrect predictions better than single-model estimates.

H4 (RQ4): Single-model uncertainty estimation methods (PriorNet, ICE with probability-based NCM) can offer lightweight alternatives but may sacrifice accuracy compared to ensemble-based approaches.

If these hypotheses hold true, they suggest that integrating ensemble-based uncertainty estimates with conformal prediction can provide a robust decision-making framework for malware classification in dynamic, real-world environments. Positive answers to RQ1–RQ4 imply more reliable automated malware detection systems that both reduce false positives—avoiding unnecessary quarantines of benign software—and minimize false negatives, allowing fewer malware instances to bypass defenses. Additionally, these improvements enable better handling of dataset shifts, which is critical as malware continuously evolves in the wild. Conversely, if the hypotheses are disproven, it would indicate that these methods may not offer significant advantages over simpler or traditional approaches.

These research questions guided our empirical design. RQ1 was addressed through calibration experiments (Section VI-A), RQ2 through analysis on IID datasets (Section VI-B), RQ3 through evaluations under dataset shifts using UCSB/BODMAS (Section VI-C), and RQ4 by comparing PriorNet and Ensemble-based methods in terms of performance versus resource cost.

\section{Empirical Analysis Methodology} \label{Emprical_Methodology}
A comprehensive approach is essential to effectively tackle the challenges of accurately accepting correct predictions and rejecting incorrect ones. Firstly, choosing a malware detector renowned for its precision in identifying malware across various datasets is crucial. Following this selection, the next step involves estimating the uncertainty associated with the detector's predictions. This process is critical for understanding the level of confidence in each prediction. With the uncertainty estimated, determining an appropriate threshold is necessary, and it is done using a validation set, which serves as a benchmark for decision-making. Once the threshold is set, the final step is to evaluate the detector's performance on a separate test set.

\subsection{Dataset description}
Our preliminary analysis is centred on the classification task using the PE malware dataset, explicitly focusing on the EMBER dataset 2018 \cite{anderson2018EMBER}, the UCSB Packed Malware dataset \cite{aghakhani2020malware}, and the BODMAS dataset \cite{yang2021bodmas}. These datasets were chosen to serve two primary purposes. Firstly, the BODMAS dataset, being more recent than EMBER, allows us to evaluate how our classifier and uncertainty estimation methods perform with newer data. Secondly, the UCSB dataset provides a unique challenge with its focus on packed malware and benign instances, which use packing or obfuscation to either evade detection or protect intellectual property. When the UCSB data is tested on a model trained with the EMBER dataset, a covariate shift in the data is introduced.   Specifically,  Let \(X_{unpacked}\) denote the feature space of the EMBER dataset (unpacked instances), and \(X_{packed}\) denote the feature space of the UCSB dataset (packed instances). The model is trained on \(P_{train}(X_{unpacked})\), expecting to generalize to \(P_{test}(X_{unpacked})\), but is instead tested on \(P_{test}(X_{packed})\). This shift, where \(P_{train}(X) \neq P_{test}(X)\), signifies a covariate shift because the distribution of the input features changes due to packing, despite the underlying prediction task (\(P(Y|X)\)) theoretically remaining constant. This scenario can be used to estimate the impact of thresholding the uncertainty estimates properly for accepting and rejecting predictions.
 The EMBER 2018 dataset, updated that year, consists of 750,000 benign and 800,000 malware instances, spanning January 2017 to December 2018. Malicious PE malware, like those in the EMBER dataset, traditionally execute unauthorized tasks, such as data theft, encryption for ransom, or system hijacks. The types of malicious behaviours in the dataset are diverse, including but not limited to ransomware, spyware, and worms. Their uniqueness is attributed to varied propagation methods, payload delivery, and evasion techniques. 
 
 The UCSB Packed Malware dataset focuses on the machine learning response to packed static features. It offers insights into 109,030 benign instances and 232,415 malware samples, primarily collected between January 2017 and March 2018 and some samples before 2017. Packed malware typically uses obfuscation to evade detection, and the UCSB dataset represents such advanced threats. In addition to these, we have also leveraged the more recent BODMAS dataset\cite{yang2021bodmas}, which encompasses data from August 2019 to September 2020, featuring 77,142 benign and 57,293 malware instances. Benign files in our datasets, such as those in the EMBER, UCSB and BODMAS collections, typically include legitimate software applications, utilities, and system files, which pose no harm and often provide essential functionalities to the user.

 These datasets capture diverse aspects of real-world malware evolution. EMBER represents a large historical corpus of Windows PE files across 2017–2018, providing broad coverage of malware families seen in that period. UCSB focuses on packed malware, which remains a significant real-world challenge as attackers routinely apply packing techniques to evade static analysis. BODMAS reflects more recent malware samples collected between 2019–2020, including newly emerging families and variants. This progression across datasets enables us to simulate temporal dataset shifts as would occur in operational malware detection pipelines. Such shifts closely resemble scenarios described in [Arp et al. 2019] and [Pendlebury et al. 2019], where classifiers face the challenge of generalising from historical to newly emerging threats.

 We acknowledge that temporally ordered dataset splits are widely recommended to prevent information leakage from future samples into the training process [Pendlebury et al. 2019; Arp et al. 2022]. In our current study, we applied random splits within EMBER and UCSB due to the lack of precise timestamp annotations in some datasets. However, to partially address this limitation, we evaluated model robustness across multiple independently collected datasets representing different collection periods (e.g., EMBER → BODMAS), which allowed us to introduce meaningful dataset shifts and approximate temporal evolution. The inclusion of UCSB (focused on packed malware) and BODMAS (containing more recent malware variants) further enabled us to simulate behavioural and temporal shifts relevant to real-world deployments. While this design introduces some diversity, future work will incorporate fully temporally ordered splits as well as extend the evaluation to additional malware formats (e.g., Android APKs, Linux ELF binaries) to better assess the generalisation and robustness of the proposed uncertainty estimation framework across multiple platforms and evolving threat landscapes.

 \subsection{Feature Extraction and Pre-processing}  
The feature extraction process for our study hinges on the features provided by the EMBER dataset\cite{anderson2018EMBER}, which have been made conveniently available across the UCSB, EMBER, and BODMAS datasets. This consistency enables a seamless comparison across these datasets and fosters a more comprehensive analysis. These features are specifically designed to train machine-learning models for static malware analysis. The features are categorized into two types: parsed features and format-agnostic features. For parsed features, EMBER uses Library to Instrument Executable Formats as a PE parser to extract general file information, header information, imported functions, exported functions, and section information. Format-agnostic features do not require parsing of the PE file, and these groups include raw byte histogram, byte entropy histogram, and string extraction. Each sample is represented as a 2381 feature vector in all three datasets. Algorithms such as neural networks are susceptible to the scale of the features, which is not the case for tree-based algorithms such as Random Forrest. To solve this, we also scale the feature to have 0 mean and standard deviation of 1 for the NN algorithm. To ensure there is no data leakage, we first split the dataset into training and testing subsets. The mean and standard deviation needed for scaling are calculated using the training data only. These parameters are then used to standardize both the training and testing datasets. 

\subsection{Classifier Selection}

In our study, we deployed four distinct classifiers to tackle the challenge of malware detection: Light Gradient Boosting Machine (LGBM), Neural Network, PriorNet, and an Ensemble of Randomly Initialized Neural Networks. The selection of LGBM was based on its prevalent use and documented success in related literature, particularly with the EMBER and BODMAS datasets \cite{anderson2018EMBER}\cite{yang2021bodmas}, where it has consistently shown high performance. This establishes LGBM as a crucial benchmark for our analysis. Neural Networks were chosen for their exceptional ability to recognize complex patterns, making them an invaluable tool for our investigation.

Furthermore, we incorporated PriorNet and an Ensemble of Neural Networks into our classifier lineup, specifically for their capabilities in uncertainty estimation. These methods are pivotal for evaluating the confidence in the predictions made by our models. In addition to these individual classifiers, we explored combinations such as LGBM+NN, LGBM+PriorNet, and LGBM+ NN-Ensemble, aiming to enhance classifier performance by leveraging the strengths of multiple models. To comprehensively assess the effectiveness of these classifiers, we employed a suite of evaluation metrics: accuracy, recall, precision, and the F1 score. 

We focused on LightGBM and Neural Networks since they represent state-of-the-art baselines in static PE malware detection [10,9], and allow implementation of uncertainty estimation techniques like ensembles, PriorNet, and conformal evaluation. While expanding to other classifiers (e.g., SVM, CNNs, GNNs) could provide broader benchmarking, our primary objective was to compare uncertainty estimation frameworks rather than classifier architectures.

\subsection{Uncertainty Estimation Techniques Selection}
In conformal prediction, a Non-Conformity Measure (NCM) quantifies how unusual a prediction is relative to calibration data. Our proposed NCMs are derived from ensemble-based uncertainty estimates (Expected Entropy, Entropy of Expected, Knowledge Uncertainty), which capture both model confidence and disagreement among ensemble members. In order to align with our research questions, we selected the following Uncertainty estimation techniques:

\textbf{Uncalibrated Probability and Calibrated Probability Estimate}: We examine the performance differential between uncalibrated and calibrated probabilities to establish a foundational baseline for our investigation. This comparison is pivotal, as uncertainty metrics like Expected Entropy and Entropy of Expected are derived from the probabilities output by the classifier. Utilizing the probability estimate from the classifier as a basis for accepting or rejecting predictions is a logical initial step, positioning it as a fundamental benchmark against all other methods evaluated in our study. 

To refine the accuracy of our calibrated probabilities and bolster the credibility of our uncertainty estimates, we employ a validation set dedicated to the calibration of our primary classifiers: Light Gradient Boosting Machine (LGBM), Neural Networks (NNs), including specialized models like Prior Net and an Ensemble of NNs. Specifically, we apply distinct calibration techniques tailored to each model type—Isotonic Regression for LGBM and Temperature Scaling for NNs—using validation data to guide the adjustment process.

Isotonic Regression is designed to recalibrate the predicted probabilities from LGBM. It achieves this by mapping the model's raw prediction scores to calibrated probabilities through a non-decreasing piecewise function. Mathematically, for a series of raw predictions \(X = \{x_1, x_2, ..., x_n\}\) and their actual outcomes \(Y = \{y_1, y_2, ..., y_n\}\), isotonic regression seeks to minimize the sum of squared differences between the observed outcomes and the function's predictions \(\min_{f}\sum_{i=1}^{n}(y_i - f(x_i))^2\), constrained by \(f(x_i) \leq f(x_{i+1})\) for each \(i\), where \(f\) represents the isotonic function.

In contrast, Temperature Scaling adjusts the prediction confidence levels for Neural Networks, PriorNet, and the NN Ensemble. This method modifies the model's logits, \(z\), by dividing them by a temperature parameter, \(T\), which can either increase (soften) or decrease (sharpen) the softmax output's decisiveness. The calibration involves identifying the optimal \(T\) that minimizes the negative log-likelihood (NLL) on the validation set. Formally, for logits \(\{z_1, z_2, ..., z_n\}\) and corresponding labels \(\{y_1, y_2, ..., y_n\}\), the goal is to solve \(\min_T -\sum_{i=1}^{n} \log\left(\frac{e^{z_{y_i}/T}}{\sum_j e^{z_j/T}}\right)\), with \(z_{y_i}\) denoting the logit for the true class \(y_i\). This process utilizes a scalar optimizer to pinpoint the ideal \(T\).

Through these calibration techniques, we aim to enhance the reliability of probability estimates and rigorously test the effectiveness of our models' uncertainty estimates in improving prediction acceptance and rejection criteria.

\textbf{Ensemble of Neural Network}: It is simply an ensemble of multiple randomly initialized neural networks. The Ensemble $f:X\to Y $ with member $f_i:X\to Y$ for $i\in1,2,....M$ can obtain the prediction by simply averaging over the member prediction,
\begin{equation}
    f(x):=\frac{1}{M}\sum_{i}^{M}f_i(x)
\end{equation}

Following the aggregation of predictions from the Ensemble, we compute key uncertainty metrics: Expected Probability, Expected Entropy, and Entropy of Expected, alongside Knowledge Uncertainty. These metrics enable the quantification of the uncertainty inherent in the Ensemble's predictions.

The Expected Probability of a given sample is calculated as the mean of the probabilities provided by each model within the Ensemble:

\begin{equation}
 E[p(x)] = \frac{1}{M} \sum_{i=1}^{M} p_i(x)   
\end{equation}

Expected Entropy reflects the average uncertainty across all individual models in the Ensemble regarding their predictions. It is a measure of the average uncertainty or the Entropy across different class predictions by the Ensemble, where a higher value suggests a broader dispersion of probability distributions among classes:

\begin{equation}\label{eqn:uncert-EE-Ensemble}
E[H(p(x))] = \frac{1}{M} \sum_{i=1}^{M} -\sum_{j} p_{ij}(x) \log(p_{ij}(x) + \epsilon) 
\end{equation}

Conversely, the Entropy of the Expected Probability assesses the uncertainty of the Ensemble's aggregated prediction, portraying the uncertainty level of the collective output:

\begin{equation}\label{eqn:uncert-EoE-Ensemble}
    H(E[p(x)]) = -\sum_{j} E[p_j(x)] \log(E[p_j(x)] + \epsilon)
\end{equation}

Finally, Knowledge Uncertainty emerges from the divergence among ensemble predictions, calculated as the difference between the Entropy of the Expected Probability and the Expected Entropy:

\begin{equation}\label{eqn:uncert-KU-Ensemble}
   Knowledge Uncertainty = H(E[p(x)]) - E[H(p(x))]
\end{equation}

\textbf{Prior Networks} aim to model the prediction uncertainty by parameterizing a prior distribution over output distributions. They emulate an ensemble of models to generate uncertainty measures but do so more efficiently, in a controlled manner, using a single network. This is achieved by parameterizing a Dirichlet distribution, which allows for the explicit representation of predictive uncertainty. The Prior Network defines a distribution over the output distributions as \(p(\pi | x^*; \hat{\theta})\), where \(x^*\) is the input, \(\pi\) represents the parameters of a categorical distribution, and
 \(\hat{\theta}\) are the parameters of the network. The network explicitly parameterizes a Dirichlet distribution as:
\begin{equation}
  p(\pi | x^*; \hat{\theta}) = p(\pi | \hat{\alpha})\quad \text{where}\quad \hat{\alpha} = f(x^*; \hat{\theta})  
\end{equation}

The Dirichlet distribution, a conjugate prior to the categorical distribution, is defined by its concentration parameters \(\alpha\), where the predictive distribution under this Dirichlet prior is given by:

\[P(y = c | x^*; \hat{\theta}) = \frac{\hat{\alpha}_c}{\sum_{k=1}^{K} \hat{\alpha}_k} = \frac{e^{\hat{z}_c}}{\sum_{k=1}^{K} e^{\hat{z}_k}}\]

This formulation captures the network's ability to model both the prediction (via the categorical distribution) and the uncertainty (via the Dirichlet distribution parameters).

Training Prior Networks involves utilizing both in-domain and out-of-distribution data. The training involves optimizing a loss function considering the reverse KL divergence between the model's output distribution and the target Dirichlet distribution for in-domain and out-of-distribution data.

For in-domain data, the loss is defined using the reverse KL-divergence between the predicted Dirichlet distribution and a target Dirichlet distribution specific to the correct class. The mathematical expression for the in-domain loss,
 \(L_{RKL}^{in}\), is:

\[L_{RKL}^{in}(y, x, \theta; \alpha^{(in)}) = \sum_{c=1}^{K} I(y = c) \cdot KL[p(\pi|x; \theta)||p(\pi|\alpha^{(in, c)})]\]

Where: \(y\) is the true class label, \(x\) is the input, \(\theta\) represents the model parameters, \(p(\pi|x; \theta)\) is the predicted Dirichlet distribution, \(p(\pi|\alpha^{(in, c)})\) is the target Dirichlet distribution for class \(c\), \(\alpha^{(in, c)}\) are the target concentration parameters for in-domain data, \(I(y = c)\) is an indicator function that equals 1 if \(y = c\) and 0 otherwise.

For out-of-distribution data, the loss function, \(L_{RKL}^{out}\), aims to train the model to predict a uniform or broad Dirichlet distribution, indicating high uncertainty for inputs that do not belong to the training data distribution:
\begin{equation}
    L_{RKL}^{out}(x, \theta; \alpha^{(out)}) = KL[p(\pi|x; \theta)||p(\pi|\alpha^{(out)})
\end{equation}

Where: \(p(\pi|\alpha^{(out)})\) is the target Dirichlet distribution for out-of-distribution data, \(\alpha^{(out)}\) is the target concentration parameters set to encourage the model to predict a uniform distribution across classes for OOD inputs.

The overall loss function combines these two components, often with a balancing parameter \(\lambda\) to control the influence of the out-of-distribution data on the training process:
\begin{equation}
L(\theta; D^{in}, D^{out}, \alpha^{(in)}, \alpha^{(out)}, \lambda) = L_{RKL}^{in} + \lambda \cdot L_{RKL}^{out}
\end{equation}

One of the challenges in training Prior Network is the requirement of OOD data. In domains like malware detection, OOD samples are not just any samples outside the training distribution but should be meaningfully representative of potential future threats or benign behaviours not covered by the training data. This specificity makes finding or generating such data difficult. Moreover, Malware and cybersecurity threats evolve rapidly, meaning what is considered OOD today might not be OOD tomorrow. To address these challenges, we utilize the Carlini \& Wagner (C\&W) attack ~\cite{carlini2017towards} 
on a neural network that has been previously trained on existing malware data. The C\&W attack method perturbs input samples to generate adversarial examples that are designed to be misclassified by the neural network.
Mathematically, this involves solving an optimization problem that minimizes the perturbation added to the input while ensuring that the perturbed example is misclassified
The objective function can be represented as:

\begin{equation}
\text{minimize } \| \delta \|_2 + c \cdot f(x + \delta, t)
\end{equation}

where \(x\) is the original input, \(\delta\) is the perturbation, \(c\) is a constant that balances the two terms, and \(f(\cdot, t)\) is the model's output for target class \(t\) indicating misclassification. This process effectively places the adversarial examples near the classifier's decision boundary yet distinctly outside the training distribution, making them valuable proxies for OOD data. 

The uncertainty estimates Probability, Expected Entropy, and Entropy of Expected of the Prior Net are calculated to estimate the uncertainty of the prediction. Given a set of concentration parameters \(\alpha\) obtained from a model for an input, the probabilities can be calculated as:

\begin{equation}
    \alpha_0 = \sum_{k=1}^{K} \alpha_k
\end{equation}

\begin{equation}
    P(y=k | x) = \frac{\alpha_k}{\alpha_0}, \quad k = 1, 2
\end{equation}

where \(x\) represents the input, \(P(y=k | x)\) is the Probability of class \(k\) given input \(x\), \(\alpha_k\) are the concentration parameters for each class \(k\), \(K\) is the total number of classes, \(\alpha_0\) is the sum of all concentration parameters for a given input, serving as a normalizing factor to obtain the probabilities.

The expected Entropy of the predictive distribution is calculated as:

\begin{equation}
    \text{Expected Entropy} = -\sum_{k=1}^{K} \left(\frac{\alpha_k}{\alpha_0}\right) \left(\psi(\alpha_k + 1) - \psi(\alpha_0 + 1)\right)
\end{equation}
where \(\psi\) is the digamma function, which is the first derivative of the logarithm of the gamma function.

The Entropy of the expected (mean) distribution is given by:

\begin{equation}
 \text{Entropy of Expected} = -\sum_{k=1}^{K} P(y=k | x) \log(P(y=k | x))   
\end{equation}

where \(P(y=k | x) = \frac{\alpha_k}{\alpha_0}\) is the probability of class \(k\) given input \(x\), as previously defined.

\textbf{Inductive Conformal Evaluation (ICE)}, proposed by Barbero et al.\cite{barbero2022transcending}, builds on the foundation of conformal prediction introduced by Vovk et al.  \cite{vovk2005algorithmic}. This method represents a state-of-the-art framework for intelligently accepting or rejecting predictions in malware classifiers. At the heart of ICE is the concept of Non Conformity Measure (NCM), which asses how unusual a new example is compared to a set of previous examples. This measure is defined by a function \( A \), which produces a nonconformity score \( \alpha_z \) for a new example \( z \), based on its divergence from a reference set \( B \) given by the following equation:
\begin{equation}
    \alpha_z = A(B, z)
\end{equation}
For example, in the work by Barbero et al., \cite{barbero2022transcending} while evaluating the PE malware dataset, they used Probability as the NCM.

The p-value plays a critical role in conformal prediction by quantifying how well the new example adheres to the established patterns of the training data. Given the nonconformity score \( \alpha_z \) of a new example \( z \), the p-value \( p_z \) is computed as follows:

\begin{equation}\label{eqn:pvalues}
    p_z = \frac{|\{\alpha \in S : \alpha \geq \alpha_z\}|}{|S|}
\end{equation}  

Here, \( S \) is the collection of nonconformity scores for all examples in the set \( B \). A low p-value indicates that the example is significantly different from the norm, suggesting potential anomalies or the onset of concept drift. ICE refines these concepts by structuring the calculation of p-values through a division of data into a proper training set and a calibration set. The model is trained exclusively on the proper training set, while nonconformity scores and p-values are derived from the calibration set. This division ensures that p-values are calculated without bias from the data used in model training, enhancing the reliability of the evaluation process. In the context of classification, p-values are computed in a label-conditional manner. This means that the reference set \( B \) is composed only of previous examples that belong to the same class \( \hat{y} \) as the class predicted for the new example \( z^* \), where \( \hat{y} = g(z^*) \). This approach ensures that the conformity of the new example is evaluated specifically against examples of its predicted class.

\textbf{Proposed Inductive Conformal Evaluation}, in this work, we proposed to use uncertainty estimate from NN Ensemble as described in equation \ref{eqn:uncert-EE-Ensemble}, \ref{eqn:uncert-EoE-Ensemble} and \ref{eqn:uncert-KU-Ensemble} as the NCM for ICE framework. Traditional NCMs, such as probability outputs from a single model, can often be misleadingly overconfident, especially in cases where the training data are not representative of the entire feature space or when they contain noisy labels. Neural network ensembles, through their aggregated predictions, provide a more nuanced view of uncertainty. Equations \ref{eqn:uncert-EE-Ensemble} and \ref{eqn:uncert-EoE-Ensemble} describe the expected Entropy and the Entropy of the expected predictions, respectively. These measures capture the variability in predictions across different models in the Ensemble, offering a more reliable assessment of uncertainty than single models. In dynamic environments where malware signatures and behaviours continuously evolve, single-model predictions might not provide the best uncertainty estimate. An ensemble's varied base learners can detect subtle shifts in data distributions more effectively. The knowledge uncertainty measure, defined in Equation \ref{eqn:uncert-KU-Ensemble}, quantitatively expresses the difference between the uncertainty of an average model and the average uncertainties of individual models. This metric is particularly useful in highlighting areas where the ensemble members disagree significantly on predictions, often a signal of novel or anomalous data points. These measures capture the variability in predictions across different models in the Ensemble, offering a more reliable assessment of uncertainty than single models. The above claims are shown in Answering RQ3 (Section \ref{RQ3}) using the model train on Ember dataset and tested with a packed malware dataset called UCSB dataset where the classifier suffered a huge F1 decrease to 72.16. For practical implementation, each new example $z$ calculates the nonconformity score using the uncertainty measures described above in \ref{eqn:uncert-EE-Ensemble}, \ref{eqn:uncert-EoE-Ensemble} and \ref{eqn:uncert-KU-Ensemble}. This score quantifies how much the example deviates from the typical predictions made by the Ensemble. The p-value for the new example 
$z$ is computed by comparing its nonconformity score against the scores from a calibration set using equation \ref{eqn:pvalues}. Like the original ICE, p-values are computed in a label-conditional manner. Finally, set a threshold \( \tau \) for p-values. If \( p_z \geq \tau \), accept the prediction; otherwise, reject it.

\subsection{\textbf{Proposed threshold estimation technique of uncertainty estimates to accept and reject Prediction}}
% \begin{figure}[ht]
%     \centering
%     % \begin{minipage}{0.46\textwidth}
%         \centering
%         \includegraphics[width=0.46\textwidth]{accet_reject_prob_EMBER.png}
%         \caption{Correctly Accepted and Correctly Rejected Trade off Graph using range of Probability threshold on EMBER Test dataset. Where, UC:Uncalibrated, C-Calibrated,Correctly Accepted:A, Correctly Rejected: R}
%         \label{fig:Prob-accept-reject-EMBER}
%     % \end{minipage}\hfill 
% \end{figure}
\begin{figure}[ht]
     % \begin{minipage}{0.46\textwidth}
        \centering
        \includegraphics[width=0.42\textwidth]{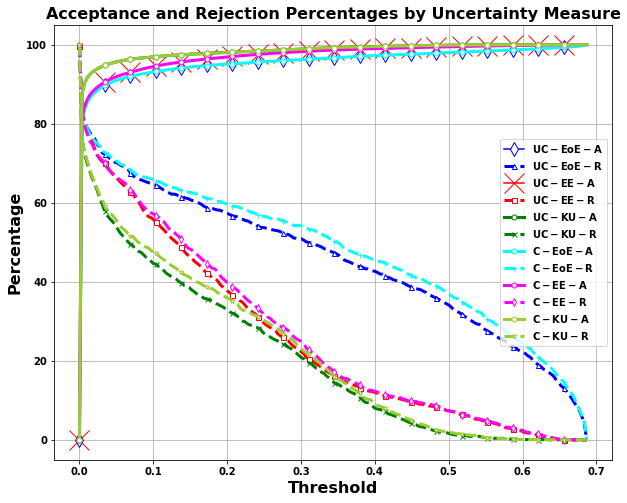}
        \caption{Correctly Accepted and Correctly Rejected Trade-off Graph using Uncertainity Estimates from NN ensemble on EMBER Test dataset. Where: UC: Uncalibrated, C-Calibrated, Correctly Accepted: A, Correctly Rejected: R, EE: Expected Entropy, EoE: Entropy of Expected and KU: Knowledge Uncertainity  }
        \label{fig:ensemble-accept-reject-EMBER}
    % \end{minipage}\hfill

\end{figure}

In the optimization strategy proposed by Barbero et al.\cite{barbero2022transcending}, the objective is to maximize a performance metric, such as the F1 score of all accepted predictions, while maintaining a rejection rate. For example, in the original Android malware classification, they used a rejection rate of less than 15\%. This method is formalized as maximizing \( F(Y, \hat{Y}, P; T) \) subject to \( G(Y, \hat{Y}, P; T) \geq C \), where \( Y \) and \( \hat{Y} \) represent vectors of ground truth and predicted labels respectively, and \( P \) is a matrix of calibration p-values, \(F\) is the performance metric and G is the constraint such as less than 15\% rejection rate. However, using hard constraints on the rejection rate may not be effective in scenarios where data variability and the nature of predictions demand flexibility. Specifically, by enforcing a rejection rate of less than 15\%, the model may either excessively restrict or insufficiently filter predictions, depending on the inherent quality and confidence levels of the predictions themselves. This rigidity could lead to scenarios where:

\begin{itemize}

\item  The model rejects too few predictions (e.g., as low as 1\%), thereby increasing the risk of accepting incorrect predictions to achieve a high F1 score. This compromises the model's integrity by potentially lowering the threshold for acceptance.
\item Conversely, the model might reject valid predictions to adhere to the constraint, especially in cases where a more significant percentage of predictions should ideally be accepted based on their quality and the confidence of the model.
    
\end{itemize}

To address these limitations, we propose an alternative approach that aims to streamline the acceptance and rejection process by optimizing the threshold used to differentiate between correct and incorrect predictions effectively. 
This approach defines four critical metrics: Correctly Accepted instances (CA), Correctly Rejected instances (CR), Incorrectly Rejected instances (IR), and Incorrectly Accepted instances (IA).
 The acceptance or rejection of a prediction based on predictive probabilities and Inductive Conformal evaluation P-values hinges on a threshold, \( \theta \), which evaluates the model's confidence level against this predefined criterion. 
The mathematical representation for each scenario is as follows:

\begin{equation}
  \text{CA}(\theta) = \sum {1}\{p_i \geq \theta \land y_i = \hat{y}_i\}  
\end{equation}

\begin{equation}
\text{CR}(\theta) = \sum{1}\{p_i < \theta \land y_i \neq \hat{y}_i\}
\end{equation}

\begin{equation}
\text{IR}(\theta) = \sum{1}\{p_i < \theta \land y_i = \hat{y}_i\}
\end{equation}

\begin{equation}
\text{IA}(\theta) = \sum{1}\{p_i \geq \theta \land y_i \neq \hat{y}_i\}
\end{equation}

For uncertainty estimates, such as expected entropy, entropy of expected, and knowledge uncertainty, the criteria for accepting or rejecting predictions adjust to whether the uncertainty measure is below a specific threshold, $\theta_u$. The formulations for decisions based on uncertainty measures are as follows:

\begin{equation}
    \text{CA}(\theta_u) = \sum{1}\{\text{uncertainty estimate}_i \leq \theta_u \land y_i = \hat{y}_i\}
\end{equation}

\begin{equation}
\text{CR}(\theta_u) = \sum {1}\{\text{uncertainty estimate}_i > \theta_u \land y_i \neq \hat{y}_i\}
\end{equation}

\begin{equation}
    \text{IR}(\theta_u) = \sum {1}\{\text{uncertainty estimate}_i > \theta_u \land y_i = \hat{y}_i\}
\end{equation}

\begin{equation}
   \text{IA}(\theta_u) = \sum {1}\{\text{uncertainty estimate}_i \leq \theta_u \land y_i \neq \hat{y}_i\} 
\end{equation}

To identify the optimal threshold for accepting or rejecting predictions based on predictive probabilities, uncertainty measures, or conformal prediction intervals, we systematically iterate over a spectrum of potential threshold values. By performing an exhaustive search, we aim to pinpoint the threshold that maximizes the harmonic mean of the Correctly Accepted percentage and Correctly Rejected percentage, denoted as \( H(\theta) \). This metric ensures a balanced consideration of both acceptance of correct predictions and rejection of incorrect ones, encapsulated by the equation:

\begin{equation}
 \text{argmax}_{\theta} \, H(\theta) = \text{argmax}_{\theta} \left( 2 \times \frac{CA\%(\theta) \times CR\%(\theta)}{CA\%(\theta) + CR\%(\theta)} \right)
\end{equation}

In this formulation, \(CA\%(\theta)\) represents the percentage of Correctly Accepted predictions, calculated as $\frac{CA(\theta)}{\sum{1}\{y_i = \hat{y}_i\}} \times 100$, while \(CR\%(\theta)\) signifies the percentage of Correctly Rejected predictions, computed as $\frac{CR(\theta)}{\sum {1}\{y_i \neq \hat{y}_i\}} \times 100$. This balanced approach aims to maximize both \( CA\%(\theta) \) and \( CR\%(\theta) \), ensuring that correct predictions are Accepted and also Correct predictions are not rejected. The harmonic mean is particularly sensitive to the values of both metrics and heavily penalizes the model if either metric is too low, helping to maintain a rigorous balance and ensuring that neither metric is sacrificed at the expense of the other.

To show that the Harmonic mean indeed provides a balance threshold of Correctly Accepted and Correctly Rejected using the validation set, we plot the trade-off between Correctly Accepted and Correctly Rejected for Uncertainty estimate from NN Ensemble, and it is shown in Figure \ref{fig:ensemble-accept-reject-EMBER}. The trade-off graphs for the rest of the technique are included in the APPENDIX.

\subsection{Integration of Methods and Role of Each Component}

Our approach does not rely on a single end-to-end pipeline. Instead, we train four classifiers independently: LightGBM (LGBM), Neural Networks (NN), PriorNet, and an Ensemble of Neural Networks. The calibrated outputs of these classifiers are then combined at the score level. This separation allows us to clearly distinguish between classifier learning, probability calibration, optional score fusion, and the decision rules for acceptance or rejection, which are implemented either through simple thresholding or via Inductive Conformal Evaluation (ICE).

Each classifier is trained on the training split and subsequently calibrated on the validation split to enhance the reliability of the probability estimates. We apply isotonic regression to LGBM, while temperature scaling is employed for the NN, PriorNet, and the Neural Network ensemble. Calibration is crucial in this context, as the effectiveness of both threshold-based decisions and non-conformity measures within ICE relies on the quality of the underlying probabilities. As shown in Table~\ref{classifier_performance}, calibration consistently improves the F1 score on in-distribution tests.

When combining models, such as LGBM+NN, LGBM+PriorNet, and LGBM+NN-Ensemble, the integration is conducted through a weighted average of the calibrated probabilities.

\begin{equation}
\tilde{p} = w \, \tilde{p}_{\text{neural}} + (1 - w) \, \tilde{p}_{\text{LGBM}}
\end{equation}

The weight $w$ is tuned on the validation set to maximize the F1 score. Equal weighting is used for the EMBER dataset, while a neural-model bias of $w=0.8$ is adopted for UCSB.

Uncertainty information is incorporated according to the design of each model. Probabilities from both LGBM and NN, whether calibrated or uncalibrated, serve as baseline signals for thresholding. The NN-Ensemble provides three complementary uncertainty measures: Expected Entropy, Entropy of Expected, and Knowledge Uncertainty. These measures can be used directly for thresholding or as non-conformity measures within ICE . In contrast, PriorNet parameterizes a Dirichlet distribution over class probabilities, producing explicit uncertainty estimates that can also be employed in thresholding or in ICE (see Table~\ref{tab:prior_net_accept_reject}). Across our experiments, ensemble-derived uncertainty measures have proven to be more effective than those from PriorNet under conditions of dataset shift.

Finally, we apply two families of decision rules—and both are thresholded. 
(i) \textit{Thresholding on scores}: we choose a probability/uncertainty threshold $\theta$ on the validation set by maximizing the harmonic mean of CA\% and CR\% (see Sec $E$ and Table ~\ref{tab:probability_accept_reject}). 
(ii) \textit{ICE}: we compute label-conditional $p$-values using either calibrated probability (baseline) or ensemble-derived uncertainty (EE, EoE, KU) as the NCM, and accept iff $p_z \ge \tau$, where the $p$-value threshold $\tau$ is selected by the \emph{same} CA/CR harmonic-mean criterion in Sec. $E$. The resulting trade-offs are reported in Tables, \ref{tab:conformal_accept_reject}, \ref{tab:ensemble_accept_reject}, \ref{tab:prior_net_accept_reject}, and \ref{tab:conformal_accept_reject_uncertainity}.

The use of ensembles further strengthens the system by improving robustness. Averaging the outputs of $M$ independently initialised neural networks reduces the variance of predicted probabilities, which mitigates overconfidence. Moreover, the disagreement among ensemble members quantified through Expected Entropy, Entropy of Expected, and Knowledge Uncertainty tends to increase under distributional shift, providing a reliable signal for rejecting uncertain predictions (Tables~\ref{tab:ensemble_accept_reject}, \ref{tab:conformal_accept_reject_uncertainity}).

\section{Experimental Results and Analysis}\label{experimnetal_result}

\subsection{Experimental Setup}
\subsubsection{Dataset Split for testing and validation}
To systematically assess the efficacy of uncertainty estimation both in the absence and presence of dataset shifts, we structured our experimental design around the EMBER, UCSB, and BODMAS datasets. Specifically, we allocated 60\% of the EMBER and UCSB datasets for training, 10\% for calibration, and 30\% for testing. This distribution aims to test the uncertainty estimates under conditions without dataset shifts rigorously. Subsequently, we employed the entire BODMAS dataset to test the model previously trained on the EMBER dataset and the uncertainty threshold derived from the same. This step is crucial for evaluating how well our uncertainty estimation performs when introduced to a newer dataset, thereby assessing its adaptability and robustness. Finally, to thoroughly understand the impact of dataset shifts on uncertainty estimation methods, we tested the UCSB Packed Malware dataset on models trained using the EMBER dataset. This approach allows us to directly observe the effectiveness of our uncertainty estimation methods in scenarios where significant shifts in dataset characteristics are present, offering insights into their reliability across varying conditions.

\subsubsection{Classifier Setup and Hyperparameters}

\begin{table*}
\centering
\caption{Classifier Performance (in \%) }
\label{classifier_performance}
\begin{tabular}{|*{9}{c|} }
    \hline
\textbf{Model}& \multicolumn{4}{c|}{\textbf{UCSB}} & \multicolumn{4}{c|}{\textbf{EMBER}}\\
    \hline
  & Accuracy  &  Precision  &   Recall  &   F1  &  Accuracy  &  Precision  &   Recall &   F1 \\
    \hline
LGBM &   97.25 &	99.37 &	96.67	& 98.00 & 99.24 & 99.11 & 99.43 & 99.27 \\
    \hline
% RF & 94.71 & 98.74 & 93.77 &	96.19	&	98.22 &	97.72 &	98.82 &	98.26 \\
%     \hline
NN  & 95.75  & 	96.50  & 	97.25  & 	96.88  & 98.63  & 	98.73  & 	98.61 & 98.67 \\
\hline
%     \hline
LGBM+NN & 97.83 & 98.81 & 98.01 & 98.41 & 99.30 &	99.19 &	99.45 & 99.32 \\
\hline
LGBM+NN+Calibrated & 97.95 &	98.42 &	98.54 &	98.48 & 99.50 &	99.53 & 99.51 & 99.52 \\
\hline
Prior-Net & 95.89 & 96.42  & 97.56 & 96.99 & 98.62 & 98.60 & 98.72 & 98.66\\ 
\hline
LGBM+Prior-Net & 97.83 & 98.82  & 97.99 & 98.40 & 98.84 & 98.79 & 98.95 & 98.87\\ 
\hline
LGBM+Prior-Net+Calibrated & 97.86 & 98.50  & 98.34 & 98.42 & 98.66 & 98.64 & 98.75 & 98.70 \\
\hline

NN-Ensemble & 95.59 & 95.92  & 97.64 & 96.77 & 98.59 & 98.91 & 98.36 & 98.63 \\ 
\hline
LGBM+NN-Ensemble & 97.81 & 98.80  & 97.98 & 98.39 & 99.47 & 99.25 & 99.72 & 99.49 \\ 
\hline
LGBM+NN-Ensemble+Calibrated & \textbf{97.97} & \textbf{98.46}  & \textbf{98.54} & \textbf{98.50} & \textbf{99.51} & \textbf{99.50} & \textbf{99.56} & \textbf{99.53} \\ 

\hline
\end{tabular}
\end{table*}

\begin{table*}[htbp]
\centering
\caption{Classifier Performance of BODMAS and UCSB dataset when tested on Model trained with EMBER (in \%) }
\label{classifier_performance_dataset_shift}
\begin{tabular}{|*{9}{c|} }
    \hline
\textbf{Model}& \multicolumn{4}{c|}{\textbf{BODMAS}} & \multicolumn{4}{c|}{\textbf{UCSB}}\\
    \hline
  & Accuracy  &  Precision  &   Recall  &   F1  &  Accuracy  &  Precision  &   Recall &   F1 \\
    \hline
LGBM+NN-Ensemble+Calibrated & \textbf{99.23} & \textbf{99.18}  & \textbf{99.02} & \textbf{99.10} & \textbf{72.16} & \textbf{98.41} & \textbf{71.22} & \textbf{72.16} \\ 

\hline
\end{tabular}
\end{table*}

Our experimental design encompassed a comprehensive array of classifiers, including LightGBM (LGBM), a Neural Network (NN), PriorNet, and an Ensemble of ten Neural Networks, alongside combinations of LGBM with each of the other classifiers. The configuration for LGBM was tailored to align with the hyperparameters used in the BODMAS setup, featuring a gradient boosting decision tree (gbdt) boosting type and a binary objective. The LGBM model was trained over 1000 iterations with a learning rate of 0.05, employing 2048 leaves, a maximum depth of 15, a minimum of 50 data points in one leaf, and a feature fraction of 0.5. As LightGBM does not yield predictions in the form of probabilities, we converted these outputs into probabilities by computing two probabilities for every prediction corresponding to the two classes. We set the probability for class 1 to match the model's prediction, while we set the probability for class 0 to be one minus the model's prediction.

The Neural Network was architectured with multiple hidden layers, specifically with sizes [1024, 1024, 1024, 500, 500, 300, 100, 200, 100, 50, 30], culminating in an output layer 2. A dropout rate of 0.2 was applied to prevent overfitting, with ReLU functioning as the activation for hidden layers and Softmax for the output layer. The architecture incorporated skip connections, bypassing alternate layers to facilitate better information flow and address potential vanishing gradient issues.

For PriorNet, we adopted a simplified hidden layer structure of [1024, 1024,30], leading to a Softplus-activated output layer of 2, suitable for predicting strictly positive $\alpha_i$ values. The loss function for PriorNet differentiates between in-domain and out-of-distribution (OOD) data: in-domain data parameters, $\alpha^{(in)}$, was set to 100 to favour sharp Dirichlet distributions denoting high prediction confidence; conversely, OOD parameters, $\alpha^{(out)}$, were adjusted to 1 to induce broad Dirichlet distributions reflective of heightened uncertainty. The training loss function maintained a balance between in-domain and adversarial example influence through a $\lambda$ value of 1.

The Ensemble comprised ten NNs, mirroring the single NN architecture but with independent random initializations and training processes. In blending LGBM with NN, PriorNet, or the NN Ensemble, we integrated the models' outputs using a weighted sum approach, assigning equal weights for the EMBER dataset and different weights (0.8 for NN, PriorNet, or NN Ensemble and 0.2 for LGBM) for the UCSB dataset, with optimal weights derived from the validation set.

During Temperature scaling to find optimal temperature the bounds are set to 0.01 and 10. The bounds are obtained using trial and error on the validation set. The general performance of the classifier such as Accuracy, Precision, Recall and F1 is shown in Table \ref{classifier_performance} where the testing is done with the split from the same dataset.

% \subsection{Answering Research Questions} \label{research_questions}
% Ideally, we would like to increase correct Prediction Acceptance and reduce Incorrect Prediction Acceptance.  

\subsection{Answering RQ1}

When using probability thresholds to decide whether to accept or reject predictions, Calibration clearly reduces the acceptance of incorrect predictions, as evidenced by the data from the EMBER and UCSB datasets shown in Table \ref{tab:probability_accept_reject}. For the EMBER dataset, the LGBM+NN model reduced incorrect acceptances (IA) from 198 to 119, LGBM+PriorNet from 261 to 170, and LGBM+NN-Ensemble from 147 to 96, showing the most significant improvement. Similarly, in the UCSB dataset, we observed reductions from 177 to 103 for LGBM+NN, from 194 to 84 for LGBM+PriorNet, and from 209 to 88 for LGBM+NN-Ensemble. However, while Calibration helps reduce incorrect predictions, it also slightly lowers the acceptance of correct predictions (CA). In the EMBER dataset, CA dropped from 291092 to 283571 for LGBM+NN, from 297249 to 294171 for LGBM+PriorNet, and from 282256 to 281026 for LGBM+NN-Ensemble. For the UCSB dataset, the decreases were from 55692 to 53081 for LGBM+NN, from 56277 to 53637 for LGBM+PriorNet, and from 53319 to 52846 for LGBM+NN-Ensemble. The model that accepted the lowest amount of incorrect predictions post-calibration was the LGBM+NN-Ensemble in the EMBER dataset, lowering the IA to 96. In the UCSB dataset, it was the LGBM+PriorNet which reduced the IA to 84. On the other hand, the uncalibrated LGBM+PriorNet model accepted the highest number of correct predictions in the EMBER dataset. This higher acceptance rate by PriorNet might be linked to its use of adversarial samples from C\&W, potentially making the classifier more confident in its decision boundary between benign and malware. In summary, using a probability threshold for accepting or rejecting predictions, Calibration tends to decrease the acceptance of incorrect predictions. 
However, this improvement comes with the trade-off of slightly reducing the acceptance of correct predictions.

 Similar to using the Probability thresholds, while using Inductive Conformal Evaluation (ICE) and Probability as the Nonconformity measure, the Calibration of Probability clearly reduces the acceptance of incorrect prediction. Post calibration of probabilities, in the Ember dataset, LGBM+NN reduced incorrect acceptances (IA) from 198 to 119, LGBM+PriorNet from 261 to 170, and LGBM+NN-Ensemble from 147 to 96, whereas in the UCSB dataset, LGBM+NN from 177 to 103, LGBM+PriorNet from 194 to 84, and LGBM+NN-Ensemble from 209 to 88. While it reduces incorrect acceptances, it also comes with a slight reduction in the acceptance of correct predictions with a decrease in CA. In the Ember dataset, the calibrated probability CA decreases in LGBM+NN from 278574 to 271178, LGBM+PriorNet from 286530 to 281630, and LGBM+NN-Ensemble from 275325 to 273289 whereas in UCSB dataset the CA decreases in LGBM+NN from 35918 to 36916, LGBM+PriorNet from 36495 to 44049, and LGBM+NN-Ensemble from 37033 to 37867. The highest CA  is obtained by Ucalibrated+LGBM+PriorNet, closely followed by Calibrated+LGBM+PriorNet in the Ember dataset, whereas in the UCSB dataset, it is Calibrated+LGBM+PriorNet. Calibrated-LGBM+PriorNet+ICE obtains the lowest IA in the Ember dataset, whereas in the UCSB dataset, it is uncalibrated+LGBM+NN+ICE.

 While using the LGBM+NN Ensemble with Uncertainty Estimation from NN Ensemble on the EMBER and UCSB datasets similar to other techniques, the Calibration of probabilities reduces the number of Incorrect prediction acceptance EMBER and UCSB datasets as evidenced by data shown in Table \ref{tab:ensemble_accept_reject}. However, unlike the decrease of CA in calibrated models in previous techniques Calibrated-Knowledge Uncertainty shows an increase in the acceptance of correct prediction from its uncalibrated counterpart from 270079 to 270107 in the Ember dataset, whereas in the UCSB dataset, all the calibrated models increased the CA as compared to the uncalibrated model. The lowest IA is obtained by Calibrated-Entropy of Expected-Uncertainty, accepting only 233 in the EMBER dataset and 256 in the UCSB dataset. In contrast, the highest CA is obtained by Calibrated-Knowledge Uncertainty, accepting 270107 in the EMBER dataset and 48398 in the UCSB dataset. A similar increase of CA in the UCSB dataset when calibrated is also observed while using uncertainty estimate from PriorNet and LGBM+PriorNet as shown in Table \ref{tab:prior_net_accept_reject}, however for the EMBER dataset, the CA decreases when using calibrated Probabilities. Similar to the patterns observed in other techniques, IA decreases when calibrated while using uncertainty estimates from PriorNet and LGBM+PriorNet. The lowest IA while using uncertainty estimate from Prior net is obtained by Calibrated-Entropy of Expected-Uncertainty accepting 443 in the EMBER dataset and 352 in the UCSB dataset. In contrast, the highest CA in Ember is obtained by Uncalibrated-Knowledge Uncertainty accepting 272716 correct instances dataset whereas Calibrated-Knowledge Uncertainty accepted the highest in the UCSB dataset accepting 51182 correct instances.

 Unlike other techniques, using uncertainty estimate from Ensemble as a nonconformity measure in Inductive conformal evaluation shows an increase in CA and a decrease in IA for all the calibrated models in both the datasets with one exception, and that is Cal+LGBM+NN-Ensemble+ICE which observes a decrease in CA when calibrated. The lowest IA and highest CA are obtained by Calibrated LGBM+NN-Ensemble+ICE while using Knowledge Uncertainty as a nonconformity measure in the Ember dataset. In contrast, in the UCSB dataset, Cal+LGBM+NN-Ensemble+ICE with Entropy of Expected obtained the highest CA, while Uncal+LGBM+NN-Ensemble+ICE obtained the lowest CA.

Among all the techniques, the lowest IA is obtained by using  Inductive conformal Evaluation and probability as a nonconformity measure in Calibrated-LGBM+PriorNet+ICE accepting 89 incorrect predictions closely followed by Calibrated-LGBM+NN-Ensemble using probability threshold accepting 96 incorrect predictions whereas the highest CA is obtained by Uncalibrated-LGBM+PriorNet accepting 297249 instances closely followed by Calibrated-LGBM+PriorNet accepting  294171 correct instances in Ember dataset. In the UCSB dataset, the lowest IA is obtained by Calibrated-LGBM+PriorNet with a probability threshold accepting 84 incorrect predictions, closely followed by Calibrated-LGBM+NN-Ensemble with a probability threshold accepting 88 incorrect predictions. The Uncalibrated-LGBM+PriorNet accepts the highest CA for the UCSB dataset, accepting 56277 correct instances.

\textbf{\textit{Operational Recommendation 1}}:  Applying Isotonic Regression to LGBM and Temperature Scaling to Neural Networks and PriorNet improves model F1 score as shown in Table \ref{classifier_performance}. Given the consistent reduction in the acceptance of incorrect predictions observed across both datasets upon Calibration, it is strongly recommended that calibrated probability models be implemented as a standard practice. This approach significantly minimizes the risk of accepting incorrect predictions—an essential feature in security applications where the cost of misclassification is notably high. While uncalibrated models tend to accept more correct predictions, they also pose a higher risk of accepting incorrect ones. 

\begin{table}[htbp]
\centering
\scriptsize
\caption{Performance of LGBM+NN, LGBM+Prior Net, LGBM+NN-Ensemble using Probability Threshold on EMBER and UCSB dataset}
\label{tab:probability_accept_reject}
\begin{tabulary}{\linewidth}{|L|L|L|L|L|L|L|L|L|}
\hline
\textbf{Data} & \multicolumn{8}{c|}{\textbf{EMBER dataset}} \\ \hline
\textbf{Type} & \multicolumn{4}{c|}{\textbf{Correctly Predicted}} & \multicolumn{4}{c|}{\textbf{Incorrectly Predicted}} \\ \hline
 & Inst & CA & IR & T & Inst. & IA & CR & T \\ \hline
Model &\multicolumn{8}{|c|}{Uncalibrated-LGBM+NN} \\ \hline
B & 149150 & 141776 & 7374 & 149150 & 874 & 20 & 854 & 874 \\ \hline
M & 158692 & 149316 & 9376 & 158692 & 1284 & 178 & 1106 & 1284 \\ \hline
T & 307842 & 291092 & 16750 & 307842 & 2158 & 198 & 1960 & 2158 \\ \hline
Model &\multicolumn{8}{|c|}{Calibrated-LGBM+NN} \\ \hline
B & 149245 & 134096 & 15149 & 149245 & 779 & 32 & 747 & 779 \\ \hline
M & 159231 & 149475 & 9756 & 159231 & 745 & 87 & 658 & 745 \\ \hline
T & 308476 & 283571 & 24905 & 308476 & 1524 & 119 & 1405 & 1524 \\ \hline
Model &\multicolumn{8}{|c|}{Uncalibrated-LGBM+PriorNet} \\ \hline
B & 148356 & 145958 & 2398 & 148356 & 1668 & 37 & 1631 & 1668 \\ \hline
M & 158054 & 151291 & 6763 & 158054 & 1922 & 224 & 1698 & 1922 \\ \hline
T & 306410 & \textbf{297249} & 9161 & 306410 & 3590 & 261 & 3329 & 3590 \\ \hline
Model &\multicolumn{8}{|c|}{Calibrated-LGBM+PriorNet} \\ \hline
B & 148036 & 141678 & 6358 & 148036 & 1988 & 53 & 1935 & 1988 \\ \hline
M & 157815 & 152493 & 5322 & 157815 & 2161 & 117 & 2044 & 2161 \\ \hline
T & 305851 & 294171 & 11680 & 305851 & 4149 & 170 & 3979 & 4149 \\ \hline
Model &\multicolumn{8}{|c|}{Uncalibrated-LGBM+NN-Ensemble} \\ \hline
B & 149592 & 136618 & 12974 & 149592 & 432 & 14 & 418 & 432 \\ \hline
M & 158786 & 145638 & 13148 & 158786 & 1190 & 133 & 1057 & 1190 \\ \hline
T & 308378 & 282256 & 26122 & 308378 & 1622 & 147 & 1475 & 1622 \\ \hline
Model &\multicolumn{8}{|c|}{Calibrated-LGBM+NN-Ensemble} \\ \hline
B & 149326 & 133013 & 16313 & 149326 & 698 & 19 & 679 & 698 \\ \hline
M & 159178 & 148013 & 11165 & 159178 & 798 & 77 & 721 & 798 \\ \hline
T & 308504 & 281026 & 27478 & 308504 & 1496 & \textbf{96} & 1400 & 1496 \\ \hline
\textbf{Data} & \multicolumn{8}{c|}{\textbf{UCSB dataset}} \\ \hline
\textbf{Type} & \multicolumn{4}{c|}{\textbf{Correctly Predicted}} & \multicolumn{4}{c|}{\textbf{Incorrectly Predicted}} \\ \hline
 & Inst & CA & IR & T & Inst. & IA & CR & T \\ \hline
Model &\multicolumn{8}{|c|}{LGBM+NN-Uncalibrated} \\ \hline
B & 19267 & 16608 & 2659 & 19267 & 847 & 113 & 734 & 847 \\ \hline
M & 41728 & 39084 & 2644 & 41728 & 501 & 64 & 437 & 501 \\ \hline
T & 60995 & 55692 & 5303 & 60995 & 1348 & 177 & 1171 & 1348 \\ \hline
Model &\multicolumn{8}{|c|}{Cal-LGBM+NN} \\ \hline
B & 19501 & 15712 & 3789 & 19501 & 613 & 63 & 550 & 613 \\ \hline
M & 41564 & 37369 & 4195 & 41564 & 665 & 40 & 625 & 665 \\ \hline
T & 61065 & 53081 & 7984 & 61065 & 1278 & 103 & 1175 & 1278 \\ \hline
Model &\multicolumn{8}{|c|}{Uncalibrated-LGBM+PriorNet} \\ \hline
B & 19257 & 16759 & 2498 & 19257 & 857 & 128 & 729 & 857 \\ \hline
M & 41729 & 39518 & 2211 & 41729 & 500 & 66 & 434 & 500 \\ \hline
T & 60986 & \textbf{56277} & 4709 & 60986 & 1357 & 194 & 1163 & 1357 \\ \hline
Model &\multicolumn{8}{|c|}{Calibrated-LGBM+PriorNet} \\ \hline
B & 19400 & 15821 & 3579 & 19400 & 714 & 45 & 669 & 714 \\ \hline
M & 41598 & 37816 & 3782 & 41598 & 631 & 39 & 592 & 631 \\ \hline
T & 60998 & 53637 & 7361 & 60998 & 1345 & \textbf{84} & 1261 & 1345 \\ \hline
Model &\multicolumn{8}{|c|}{Uncalibrated-LGBM+NN-Ensemble} \\ \hline
B & 19254 & 15490 & 3764 & 19254 & 860 & 135 & 725 & 860 \\ \hline
M & 41726 & 37829 & 3897 & 41726 & 503 & 74 & 429 & 503 \\ \hline
T & 60980 & 53319 & 7661 & 60980 & 1363 & 209 & 1154 & 1363 \\ \hline
Model &\multicolumn{8}{|c|}{Calibrated-LGBM+NN-Ensemble} \\ \hline
B & 19501 & 15389 & 4112 & 19501 & 613 & 55 & 558 & 613 \\ \hline
M & 41579 & 37457 & 4122 & 41579 & 650 & 33 & 617 & 650 \\ \hline
T & 61080 & 52846 & 8234 & 61080 & 1263 & 88 & 1175 & 1263 \\ \hline
\end{tabulary}
\begin{flushleft}
Legend: Inst - Number of instances; CA - Correctly Accepted; CR - Correctly Rejected; IR - Incorrectly Rejected; IA - Incorrectly Accepted; B - Benign; M - Malware; T - Total number of Instances predicted correctly or incorrectly
\end{flushleft}
\end{table}

\begin{table}[htbp]
\centering
\scriptsize
\caption{Performance of Conformal Prediction on EMBER and UCSB dataset using Probability as Non-Conformity Measure (State of the Art) }
\label{tab:conformal_accept_reject}
\begin{tabulary}{\linewidth}{|L|L|L|L|L|L|L|L|L|}
\hline
\textbf{Data} & \multicolumn{8}{c|}{\textbf{EMBER dataset}} \\ \hline
\textbf{Type} & \multicolumn{4}{c|}{\textbf{Correctly Predicted}} & \multicolumn{4}{c|}{\textbf{Incorrectly Predicted}} \\ \hline
 & \textbf{Inst} & \textbf{CA} & \textbf{IR} & \textbf{T} & \textbf{Inst.} & \textbf{IA} & \textbf{CR} & \textbf{T} \\ \hline
Model &\multicolumn{8}{|c|}{Uncalibrated-LGBM+NN+ICE} \\ \hline
B & 149150 & 139695 & 9455 & 149150 & 874 & 3 & 871 & 874 \\ \hline
M & 158692 & 138879 & 19813 & 158692 & 1284 & 141 & 1143 & 1284 \\ \hline
T & 307842 & 278574 & 29268 & 307842 & 2158 & 144 & 2014 & 2158 \\ \hline
Model &\multicolumn{8}{|c|}{Calibrated-LGBM+NN+ICE} \\ \hline
B & 149245 & 135881 & 13364 & 149245 & 779 & 2 & 777 & 779 \\ \hline
M & 159231 & 135297 & 23934 & 159231 & 745 & 98 & 647 & 745 \\ \hline
T & 308476 & 271178 & 37298 & 308476 & 1524 & 100 & 1424 & 1524 \\ \hline
Model &\multicolumn{8}{|c|}{Uncalibrated-LGBM+PriorNet+ICE} \\ \hline
B & 148039 & 143554 & 4485 & 148039 & 1985 & 6 & 1979 & 1985 \\ \hline
M & 157930 & 142976 & 14954 & 157930 & 2046 & 159 & 1887 & 2046 \\ \hline
T & 305969 & \textbf{286530} & 19439 & 305969 & 4031 & 165 & 3866 & 4031 \\ \hline
Model &\multicolumn{8}{|c|}{Calibrated-LGBM+PriorNet+ICE} \\ \hline
B & 148621 & 137735 & 10886 & 148621 & 1403 & 10 & 1393 & 1403 \\ \hline
M & 159152 & 143895 & 15257 & 159152 & 824 & 79 & 745 & 824 \\ \hline
T & 307773 & 281630 & 26143 & 307773 & 2227 & \textbf{89} & 2138 & 2227 \\ \hline
Model &\multicolumn{8}{|c|}{Uncalibrated-LGBM+NN-Ensemble+ICE} \\ \hline
B & 149592 & 137928 & 11664 & 149592 & 432 & 4 & 428 & 432 \\ \hline
M & 158786 & 137397 & 21389 & 158786 & 1190 & 141 & 1049 & 1190 \\ \hline
T & 308378 & 275325 & 33053 & 308378 & 1622 & 145 & 1477 & 1622 \\ \hline

Model &\multicolumn{8}{|c|}{Calibrated-LGBM+NN-Ensemble+ICE} \\ \hline
B & 149326 & 136987 & 12339 & 149326 & 698 & 2 & 696 & 698 \\ \hline
M & 159178 & 136302 & 22876 & 159178 & 798 & 104 & 694 & 798 \\ \hline
T & 308504 & 273289 & 35215 & 308504 & 1496 & 106 & 1390 & 1496 \\ \hline

\textbf{Data} & \multicolumn{8}{c|}{\textbf{UCSB dataset}} \\ \hline
\textbf{Type} & \multicolumn{4}{c|}{\textbf{Correctly Predicted}} & \multicolumn{4}{c|}{\textbf{Incorrectly Predicted}} \\ \hline
 & \textbf{Inst} & \textbf{CA} & \textbf{IR} & \textbf{Total} & \textbf{Inst.} & \textbf{IA} & \textbf{CR} & \textbf{T} \\ \hline
Model &\multicolumn{8}{|c|}{Uncal+LGBM+NN+ICE} \\ \hline
B & 19267 & 17516 & 1751 & 19267 & 847 & 0 & 847 & 847 \\ \hline
M & 41728 & 18402 & 23326 & 41728 & 501 & 124 & 377 & 501 \\ \hline
T & 60995 & 35918 & 25077 & 60995 & 1348 & \textbf{124} & 1224 & 1348 \\ \hline
Model &\multicolumn{8}{|c|}{Cal+LGBM+NN+ICE} \\ \hline
B & 19501 & 18013 & 1488 & 19501 & 613 & 1 & 612 & 613 \\ \hline
M & 41564 & 18903 & 22661 & 41564 & 665 & 166 & 499 & 665 \\ \hline
T & 61065 & 36916 & 24149 & 61065 & 1278 & 167 & 1111 & 1278 \\ \hline
Model &\multicolumn{8}{|c|}{Uncal+LGBM+Priornet+ICE} \\ \hline
B & 19255 & 17769 & 1486 & 19255 & 859 & 0 & 859 & 859 \\ \hline
M & 41729 & 18726 & 23003 & 41729 & 500 & 156 & 344 & 500 \\ \hline
T & 60984 & 36495 & 24489 & 60984 & 1359 & 156 & 1203 & 1359 \\ \hline
Model &\multicolumn{8}{|c|}{Cal+LGBM+Priornet+ICE} \\ \hline
B & 19397 & 18854 & 543 & 19397 & 717 & 2 & 715 & 717 \\ \hline
M & 41595 & 25195 & 16400 & 41595 & 634 & 336 & 298 & 634 \\ \hline
T & 60992 & \textbf{44049} & 16943 & 60992 & 1351 & 338 & 1013 & 1351 \\ \hline
Model &\multicolumn{8}{|c|}{Uncal+LGBM+NN-Ensemble+ICE} \\ \hline
B & 19254 & 18031 & 1223 & 19254 & 860 & 6 & 854 & 860 \\ \hline
M & 41726 & 19002 & 22724 & 41726 & 503 & 194 & 309 & 503 \\ \hline
T & 60980 & 37033 & 23947 & 60980 & 1363 & 200 & 1163 & 1363 \\ \hline
Model &\multicolumn{8}{|c|}{Cal+LGBM+NN-Ensemble+ICE} \\ \hline
B & 19501 & 18198 & 1303 & 19501 & 613 & 1 & 612 & 613 \\ \hline
M & 41579 & 19669 & 21910 & 41579 & 650 & 197 & 453 & 650 \\ \hline
T & 61080 & 37867 & 23213 & 61080 & 1263 & 198 & 1065 & 1263 \\ \hline

\end{tabulary}
\begin{flushleft}
Legend: Inst - Number of instances; CA - Correctly Accepted; CR - Correctly Rejected; IR - Incorrectly Rejected; IA - Incorrectly Accepted; B - Benign; M - Malware; T - Total number of Instances predicted correctly or incorrectly
\end{flushleft}
\end{table}

\begin{table}[htbp]
\centering
\scriptsize
\caption{Performance of LGBM+NN Ensemble with Uncertainty Estimation from NN Ensemble on EMBER and UCSB}
\label{tab:ensemble_accept_reject}
\begin{tabulary}{\linewidth}{|L|L|L|L|L|L|L|L|L|}
\hline
\textbf{Data} & \multicolumn{8}{c|}{\textbf{EMBER dataset}} \\ \hline
\textbf{Type} & \multicolumn{4}{c|}{\textbf{Correctly Predicted}} & \multicolumn{4}{c|}{\textbf{Incorrectly Predicted}} \\ \hline
 & \textbf{Inst} & \textbf{CA} & \textbf{IR} & \textbf{T} & \textbf{Inst.} & \textbf{IA} & \textbf{CR} & \textbf{T} \\ \hline
Model &\multicolumn{8}{|c|}{Uncalibrated-Expected Entropy-Uncertainty Est.} \\ \hline
B & 150024 & 123631 & 26393 & 150024 & 432 & 89 & 343 & 432 \\ \hline
M & 159976 & 141646 & 18330 & 159976 & 1190 & 268 & 922 & 1190 \\ \hline
T & 310000 & 265277 & 44723 & 310000 & 1622 & 357 & 1265 & 1622 \\ \hline
Model &\multicolumn{8}{|c|}{Calibrated-Expected Entropy-Uncertainty Est.} \\ \hline
B & 149326 & 114414 & 34912 & 149326 & 698 & 99 & 599 & 698 \\ \hline
M & 159178 & 136317 & 22861 & 159178 & 798 & 145 & 653 & 798 \\ \hline
T & 308504 & 250731 & 57773 & 308504 & 1496 & 244 & 1252 & 1496 \\ \hline
Model &\multicolumn{8}{|c|}{Uncalibrated-Entropy of Expected-Uncertainty Est.} \\ \hline
B & 149592 & 122309 & 27283 & 149592 & 432 & 85 & 347 & 432 \\ \hline
M & 158786 & 140590 & 18196 & 158786 & 1190 & 256 & 934 & 1190 \\ \hline
T & 308378 & 262899 & 45479 & 308378 & 1622 & 341 & 1281 & 1622 \\ \hline
Model &\multicolumn{8}{|c|}{Calibrated-Entropy of Expected-Uncertainty Est.} \\ \hline
B & 149326 & 113061 & 36265 & 149326 & 698 & 92 & 606 & 698 \\ \hline
M & 159178 & 135621 & 23557 & 159178 & 798 & 141 & 657 & 798 \\ \hline
T & 308504 & 248682 & 59822 & 308504 & 1496 & \textbf{233} & 1263 & 1496 \\ \hline
Model &\multicolumn{8}{|c|}{Uncalibrated-Knowledge Uncertainty} \\ \hline
B & 149592 & 127244 & 22348 & 149592 & 432 & 93 & 339 & 432 \\ \hline
M & 158786 & 142835 & 15951 & 158786 & 1190 & 291 & 899 & 1190 \\ \hline
T & 308378 & 270079 & 38299 & 308378 & 1622 & 384 & 1238 & 1622 \\ \hline
Model &\multicolumn{8}{|c|}{Calibrated-Knowledge Uncertainty} \\ \hline
B & 149326 & 127203 & 22123 & 149326 & 698 & 134 & 564 & 698 \\ \hline
M & 159178 & 142904 & 16274 & 159178 & 798 & 222 & 576 & 798 \\ \hline
T & 308504 & \textbf{270107} & 38397 & 308504 & 1496 & 356 & 1140 & 1496 \\ \hline
\textbf{Data} & \multicolumn{8}{c|}{\textbf{UCSB dataset}} \\ \hline
\textbf{Type} & \multicolumn{4}{c|}{\textbf{Correctly Predicted}} & \multicolumn{4}{c|}{\textbf{Incorrectly Predicted}} \\ \hline
 & \textbf{Inst} & \textbf{CA} & \textbf{IR} & \textbf{Total} & \textbf{Inst.} & \textbf{IA} & \textbf{CR} & \textbf{T} \\ \hline
Model &\multicolumn{8}{|c|}{Uncalibrated-Expected Entropy Est.} \\ \hline
B & 19254 & 13506 & 5748 & 19254 & 860 & 196 & 664 & 860 \\ \hline
M & 41726 & 32352 & 9374 & 41726 & 503 & 84 & 419 & 503 \\ \hline
T & 60980 & 45858 & 15122 & 60980 & 1363 & 280 & 1083 & 1363 \\ \hline
Model &\multicolumn{8}{|c|}{Calibrated-Expected Entropy Est.} \\ \hline
B & 19501 & 13527 & 5974 & 19501 & 613 & 175 & 438 & 613 \\ \hline
M & 41579 & 32344 & 9235 & 41579 & 650 & 92 & 558 & 650 \\ \hline
T & 61080 & 45871 & 15209 & 61080 & 1263 & 267 & 996 & 1263 \\ \hline
Model &\multicolumn{8}{|c|}{Uncalibrated-Entropy of Expected Est.} \\ \hline
B & 19254 & 13209 & 6045 & 19254 & 860 & 190 & 670 & 860 \\ \hline
M & 41726 & 32365 & 9361 & 41726 & 503 & 78 & 425 & 503 \\ \hline
T & 60980 & 45574 & 15406 & 60980 & 1363 & 268 & 1095 & 1363 \\ \hline
Model &\multicolumn{8}{|c|}{Calibrated-Entropy of Expected Est.} \\ \hline
B & 19501 & 13230 & 6271 & 19501 & 613 & 169 & 444 & 613 \\ \hline
M & 41579 & 32356 & 9223 & 41579 & 650 & 87 & 563 & 650 \\ \hline
T & 61080 & 45586 & 15494 & 61080 & 1263 & \textbf{256} & 1007 & 1263 \\ \hline
Model &\multicolumn{8}{|c|}{Uncalibrated-Knowledge Uncertainty Est.} \\ \hline
B & 19254 & 13831 & 5423 & 19254 & 860 & 283 & 577 & 860 \\ \hline
M & 41726 & 34539 & 7187 & 41726 & 503 & 127 & 376 & 503 \\ \hline
T & 60980 & 48370 & 12610 & 60980 & 1363 & 410 & 953 & 1363 \\ \hline
Model &\multicolumn{8}{|c|}{Calibrated-Knowledge Uncertainty Est.} \\ \hline
B & 19501 & 13879 & 5622 & 19501 & 613 & 235 & 378 & 613 \\ \hline
M & 41579 & 34519 & 7060 & 41579 & 650 & 147 & 503 & 650 \\ \hline
T & 61080 & \textbf{48398} & 12682 & 61080 & 1263 & 382 & 881 & 1263 \\ \hline
\end{tabulary}
\begin{flushleft}
Legend: Inst - Number of instances; CA - Correctly Accepted; CR - Correctly Rejected; IR - Incorrectly Rejected; IA - Incorrectly Accepted; B - Benign; M - Malware; T - Total number of Instances predicted correctly or incorrectly
\end{flushleft}
\end{table}

\begin{table}[htbp]
\centering
\scriptsize
\caption{Performance of LGBM+Prior Net with Uncertainty Estimation from Prior Net on EMBER and UCSB dataset}
\label{tab:prior_net_accept_reject}
\begin{tabulary}{\linewidth}{|L|L|L|L|L|L|L|L|L|}
\hline
\textbf{Data} & \multicolumn{8}{c|}{\textbf{EMBER dataset}} \\ \hline
\textbf{Type} & \multicolumn{4}{c|}{\textbf{Correctly Predicted}} & \multicolumn{4}{c|}{\textbf{Incorrectly Predicted}} \\ \hline
 & \textbf{Inst} & \textbf{CA} & \textbf{IR} & \textbf{T} & \textbf{Inst.} & \textbf{IA} & \textbf{CR} & \textbf{T} \\ \hline
Model &\multicolumn{8}{|c|}{Uncalibrated-Expected Entropy-Uncertainty Est.} \\ \hline
B & 148356 & 122413 & 25943 & 148356 & 1668 & 254 & 1414 & 1668 \\ \hline
M & 158054 & 142587 & 15467 & 158054 & 1922 & 360 & 1562 & 1922 \\ \hline
T & 306410 & 265000 & 41410 & 306410 & 3590 & 614 & 2976 & 3590 \\ \hline
Model &\multicolumn{8}{|c|}{Calibrated-Expected Entropy-Uncertainty Est.} \\ \hline
B & 148949 & 112349 & 36600 & 148949 & 1075 & 199 & 876 & 1075 \\ \hline
M & 159030 & 136544 & 22486 & 159030 & 946 & 271 & 675 & 946 \\ \hline
T & 307979 & 248893 & 59086 & 307979 & 2021 & 470 & 1551 & 2021 \\ \hline
Model &\multicolumn{8}{|c|}{Uncalibrated-Entropy of Expected-Uncertainty Est.} \\ \hline
B & 148356 & 119909 & 28447 & 148356 & 1668 & 243 & 1425 & 1668 \\ \hline
M & 158054 & 141127 & 16927 & 158054 & 1922 & 325 & 1597 & 1922 \\ \hline
T & 306410 & 261036 & 45374 & 306410 & 3590 & 568 & 3022 & 3590 \\ \hline
Model &\multicolumn{8}{|c|}{Calibrated-Entropy of Expected-Uncertainty Est.} \\ \hline
B & 148949 & 109528 & 39421 & 148949 & 1075 & 186 & 889 & 1075 \\ \hline
M & 159030 & 134892 & 24138 & 159030 & 946 & 257 & 689 & 946 \\ \hline
T & 307979 & 244420 & 63559 & 307979 & 2021 & \textbf{443} & 1578 & 2021 \\ \hline
Model &\multicolumn{8}{|c|}{Uncalibrated-Knowledge Uncertainty Est.} \\ \hline
B & 148356 & 126738 & 21618 & 148356 & 1668 & 319 & 1349 & 1668 \\ \hline
M & 158054 & 145978 & 12076 & 158054 & 1922 & 425 & 1497 & 1922 \\ \hline
T & 306410 & \textbf{272716} & 33694 & 306410 & 3590 & 744 & 2846 & 3590 \\ \hline
Model &\multicolumn{8}{|c|}{Calibrated-Knowledge Uncertainty Est.} \\ \hline
B & 148949 & 113728 & 35221 & 148949 & 1075 & 225 & 850 & 1075 \\ \hline
M & 159030 & 138022 & 21008 & 159030 & 946 & 288 & 658 & 946 \\ \hline
T & 307979 & 251750 & 56229 & 307979 & 2021 & 513 & 1508 & 2021 \\ \hline
\textbf{Data} & \multicolumn{8}{c|}{\textbf{UCSB dataset}} \\ \hline
\textbf{Type} & \multicolumn{4}{c|}{\textbf{Correctly Predicted}} & \multicolumn{4}{c|}{\textbf{Incorrectly Predicted}} \\ \hline
 & \textbf{Inst} & \textbf{CA} & \textbf{IR} & \textbf{T} & \textbf{Inst.} & \textbf{IA} & \textbf{CR} & \textbf{T} \\ \hline
Model &\multicolumn{8}{|c|}{Uncalibrated-Expected Entropy Est.} \\ \hline
B & 19261 & 16469 & 2792 & 19261 & 853 & 298 & 555 & 853 \\ \hline
M & 41731 & 32771 & 8960 & 41731 & 498 & 166 & 332 & 498 \\ \hline
T & 60992 & 49240 & 11752 & 60992 & 1351 & 464 & 887 & 1351 \\ \hline
Model &\multicolumn{8}{|c|}{Calibrated-Expected Entropy Est.} \\ \hline
B & 19416 & 16517 & 2899 & 19416 & 698 & 250 & 448 & 698 \\ \hline
M & 41599 & 32737 & 8862 & 41599 & 630 & 200 & 430 & 630 \\ \hline
T & 61015 & 49254 & 11761 & 61015 & 1328 & 450 & 878 & 1328 \\ \hline
Model &\multicolumn{8}{|c|}{Uncalibrated-Entropy of Expected Est.} \\ \hline
B & 19261 & 15854 & 3407 & 19261 & 853 & 215 & 638 & 853 \\ \hline
M & 41731 & 25166 & 16565 & 41731 & 498 & 137 & 361 & 498 \\ \hline
T & 60992 & 41020 & 19972 & 60992 & 1351 & \textbf{352} & 999 & 1351 \\ \hline
Model &\multicolumn{8}{|c|}{Calibrated-Entropy of Expected Est.} \\ \hline
B & 19416 & 16513 & 2903 & 19416 & 698 & 254 & 444 & 698 \\ \hline
M & 41599 & 33019 & 8580 & 41599 & 630 & 201 & 429 & 630 \\ \hline
T & 61015 & 49532 & 11483 & 61015 & 1328 & 455 & 873 & 1328 \\ \hline
Model &\multicolumn{8}{|c|}{Uncalibrated-Knowledge Uncertainty Est.} \\ \hline
B & 19261 & 16496 & 2765 & 19261 & 853 & 319 & 534 & 853 \\ \hline
M & 41731 & 34669 & 7062 & 41731 & 498 & 167 & 331 & 498 \\ \hline
T & 60992 & 51165 & 9827 & 60992 & 1351 & 486 & 865 & 1351 \\ \hline
Model &\multicolumn{8}{|c|}{Calibrated-Knowledge Uncertainty Est.} \\ \hline
B & 19416 & 16545 & 2871 & 19416 & 698 & 270 & 428 & 698 \\ \hline
M & 41599 & 34637 & 6962 & 41599 & 630 & 199 & 431 & 630 \\ \hline
T & 61015 & \textbf{51182} & 9833 & 61015 & 1328 & 469 & 859 & 1328 \\ \hline
\end{tabulary}
\begin{flushleft}
Legend: Inst - Number of instances; CA - Correctly Accepted; CR - Correctly Rejected; IR - Incorrectly Rejected; IA - Incorrectly Accepted; B - Benign; M - Malware; T - Total number of Instances predicted correctly or incorrectly
\end{flushleft}
\end{table}

\begin{table}[htbp]
\centering
\scriptsize
\caption{Performance of Conformal Prediction on EMBER and UCSB dataset using Uncertainity from Ensemble as Non-Conformity Measure }
\label{tab:conformal_accept_reject_uncertainity}
\begin{tabulary}{\linewidth}{|L|L|L|L|L|L|L|L|L|}
\hline
\textbf{Data} & \multicolumn{8}{c|}{\textbf{EMBER dataset}} \\ \hline
\textbf{Type} & \multicolumn{4}{c|}{\textbf{Correctly Predicted}} & \multicolumn{4}{c|}{\textbf{Incorrectly Predicted}} \\ \hline
 & \textbf{Inst} & \textbf{CA} & \textbf{IR} & \textbf{T} & \textbf{Inst.} & \textbf{IA} & \textbf{CR} & \textbf{T} \\ \hline
Model &\multicolumn{8}{|c|}{Uncal+LGBM+NN-Ensemble+ICE-NCM:Expected Entropy} \\ \hline
B & 149592 & 120639 & 28953 & 149592 & 432 & 64 & 368 & 432 \\ \hline
M & 158786 & 127807 & 30979 & 158786 & 1190 & 216 & 974 & 1190 \\ \hline
T & 308378 & 248446 & 59932 & 308378 & 1622 & 280 & 1342 & 1622 \\ \hline
Model &\multicolumn{8}{|c|}{Cal+LGBM+NN-Ensemble+ICE-NCM:Expected Entropy} \\ \hline
B & 149326 & 122393 & 26933 & 149326 & 698 & 75 & 623 & 698 \\ \hline
M & 159178 & 129853 & 29325 & 159178 & 798 & 199 & 599 & 798 \\ \hline
T & 308504 & 252246 & 56258 & 308504 & 1496 & 274 & 1222 & 1496 \\ \hline
Model &\multicolumn{8}{|c|}{Uncali+LGBM+NN-Ensemble+ICE-NCM:Entropy of Expected} \\ \hline
B & 149592 & 120730 & 28862 & 149592 & 432 & 63 & 369 & 432 \\ \hline
M & 158786 & 127984 & 30802 & 158786 & 1190 & 216 & 974 & 1190 \\ \hline
T & 308378 & 248714 & 59664 & 308378 & 1622 & 279 & 1343 & 1622 \\ \hline
Model &\multicolumn{8}{|c|}{Cali+LGBM+NN-Ensemble+ICE-NCM:Entropy of Expected} \\ \hline
B & 149326 & 122261 & 27065 & 149326 & 698 & 74 & 624 & 698 \\ \hline
M & 159178 & 129750 & 29428 & 159178 & 798 & 193 & 605 & 798 \\ \hline
T & 308504 & 252011 & 56493 & 308504 & 1496 & 267 & 1229 & 1496 \\ \hline
Model &\multicolumn{8}{|c|}{Uncali+LGBM+NN-Ensemble+ICE-NCM:Knowledge Uncertainty} \\ \hline
B & 149592 & 121443 & 28149 & 149592 & 432 & 65 & 367 & 432 \\ \hline
M & 158786 & 128816 & 29970 & 158786 & 1190 & 223 & 967 & 1190 \\ \hline
T & 308378 & 250259 & 58119 & 308378 & 1622 & 288 & 1334 & 1622 \\ \hline
Model &\multicolumn{8}{|c|}{Cali+LGBM+NN-Ensemble+ICE-NCM:Knowledge Uncertainty} \\ \hline
B & 149326 & 122502 & 26824 & 149326 & 698 & 70 & 628 & 698 \\ \hline
M & 159178 & 130083 & 29095 & 159178 & 798 & 194 & 604 & 798 \\ \hline
T & 308504 & \textbf{252585} & 55919 & 308504 & 1496 & \textbf{264} & 1232 & 1496 \\ \hline

\textbf{Data} & \multicolumn{8}{c|}{\textbf{UCSB dataset}} \\ \hline
\textbf{Type} & \multicolumn{4}{c|}{\textbf{Correctly Predicted}} & \multicolumn{4}{c|}{\textbf{Incorrectly Predicted}} \\ \hline
 & \textbf{Inst} & \textbf{CA} & \textbf{IR} & \textbf{T} & \textbf{Inst.} & \textbf{IA} & \textbf{CR} & \textbf{T} \\ \hline
Model &\multicolumn{8}{|c|}{Uncal+LGBM+NN-Ensemble+ICE-NCM:Expected Entropy} \\ \hline
B & 19254 & 13912 & 5342 & 19254 & 860 & 166 & 694 & 860 \\ \hline
M & 41726 & 29654 & 12072 & 41726 & 503 & 102 & 401 & 503 \\ \hline
T & 60980 & 43566 & 17414 & 60980 & 1363 & \textbf{268} & 1095 & 1363 \\ \hline
Model &\multicolumn{8}{|c|}{Cal+LGBM+NN-Ensemble+ICE-NCM:Expected Entropy} \\ \hline
B & 19501 & 14878 & 4623 & 19501 & 613 & 165 & 448 & 613 \\ \hline
M & 41579 & 31655 & 9924 & 41579 & 650 & 151 & 499 & 650 \\ \hline
T & 61080 & 46533 & 14547 & 61080 & 1263 & 316 & 947 & 1263 \\ \hline
Model &\multicolumn{8}{|c|}{Uncal+LGBM+NN-Ensemble+ICE-NCM:Entropy of Expected} \\ \hline
B & 19254 & 14322 & 4932 & 19254 & 860 & 176 & 684 & 860 \\ \hline
M & 41726 & 30546 & 11180 & 41726 & 503 & 114 & 389 & 503 \\ \hline
T & 60980 & 44868 & 16112 & 60980 & 1363 & 290 & 1073 & 1363 \\ \hline
Model &\multicolumn{8}{|c|}{Cal+LGBM+NN-Ensemble+ICE-NCM:Entropy of Expected} \\ \hline
B & 19501 & 14355 & 5146 & 19501 & 613 & 143 & 470 & 613 \\ \hline
M & 41579 & 30529 & 11050 & 41579 & 650 & 131 & 519 & 650 \\ \hline
T & 61080 & \textbf{44884} & 16196 & 61080 & 1263 & 274 & 989 & 1263 \\ \hline
Model &\multicolumn{8}{|c|}{Uncal+LGBM+NN-Ensemble+ICE-NCM:Knowledge Uncertainty} \\ \hline
B & 19254 & 14045 & 5209 & 19254 & 860 & 212 & 648 & 860 \\ \hline
M & 41726 & 29884 & 11842 & 41726 & 503 & 132 & 371 & 503 \\ \hline
T & 60980 & 43929 & 17051 & 60980 & 1363 & 344 & 1019 & 1363 \\ \hline
Model &\multicolumn{8}{|c|}{Cal+LGBM+NN-Ensemble+ICE-NCM:Knowledge Uncertainty} \\ \hline
B & 19501 & 13975 & 5526 & 19501 & 613 & 159 & 454 & 613 \\ \hline
M & 41579 & 29615 & 11964 & 41579 & 650 & 151 & 499 & 650 \\ \hline
T & 61080 & 43590 & 17490 & 61080 & 1263 & 310 & 953 & 1263 \\ \hline

\end{tabulary}
\begin{flushleft}
Legend: Inst - Number of instances; CA - Correctly Accepted; CR - Correctly Rejected; IR - Incorrectly Rejected; IA - Incorrectly Accepted; B - Benign; M - Malware; T - Total number of Instances predicted correctly or incorrectly
\end{flushleft}
\end{table}

\begin{table}[htbp]
\centering
\scriptsize
\caption{Performance of BODMAS dataset when tested on EMBER dataset}
\label{tab:bodmas_test_EMBER_accept_reject}
\begin{tabulary}{\linewidth}{|L|L|L|L|L|L|L|L|L|}
\hline
\textbf{Data} & \multicolumn{8}{c|}{\textbf{Probability Calibration Models}} \\ \hline
\textbf{Type} & \multicolumn{4}{c|}{\textbf{Correctly Predicted}} & \multicolumn{4}{c|}{\textbf{Incorrectly Predicted}} \\ \hline
 & \textbf{Inst} & \textbf{CA} & \textbf{IR} & \textbf{T} & \textbf{Inst.} & \textbf{IA} & \textbf{CR} & \textbf{T} \\ \hline
Model &\multicolumn{8}{|c|}{Probability-Cal-LGBM+NN} \\ \hline
B & 76160 & 56194 & 19966 & 76160 & 982 & 58 & 924 & 982 \\ \hline
M & 57015 & 51839 & 5176 & 57015 & 278 & 10 & 268 & 278 \\ \hline
T & 133175 & 108033 & 25142 & 133175 & 1260 & 68 & 1192 & 1260 \\ \hline
Model &\multicolumn{8}{|c|}{Probability-Cal-LGBM+PriorNet} \\ \hline
B & 75718 & 65951 & 9767 & 75718 & 1424 & 56 & 1368 & 1424 \\ \hline
M & 56294 & 52233 & 4061 & 56294 & 999 & 18 & 963 & 999 \\ \hline
T & 132012 & 118184 & 13828 & 132012 & 2423 & 74 & 2349 & 2423 \\ \hline
Model &\multicolumn{8}{|c|}{Probability-Cal-LGBM+NN Ensemble} \\ \hline
B & 76188 & 55235 & 20953 & 76188 & 954 & 44 & 910 & 954 \\ \hline
M & 56971 & 52024 & 4947 & 56971 & 322 & 8 & 314 & 322 \\ \hline
T & 133159 & 107259 & 25900 & 133159 & 1276 & 52 & 1224 & 1276 \\ \hline
Model &\multicolumn{8}{|c|}{Expected Entropy-Cal-LGBM+PriorNet} \\ \hline
B & 75718 & 51998 & 23720 & 75718 & 1424 & 117 & 1307 & 1424 \\ \hline
M & 56294 & 47137 & 9157 & 56294 & 999 & 31 & 968 & 999 \\ \hline
T & 132012 & 99135 & 32877 & 132012 & 2423 & 148 & 2275 & 2423 \\ \hline
Model &\multicolumn{8}{|c|}{Entropy of Expected-Cal-LGBM+PriorNet} \\ \hline
B & 75718 & 54374 & 21344 & 75718 & 1424 & 147 & 1277 & 1424 \\ \hline
M & 56294 & 49103 & 7191 & 56294 & 999 & 35 & 964 & 999 \\ \hline
T & 132012 & 103477 & 28535 & 132012 & 2423 & 182 & 2241 & 2423 \\ \hline
Model &\multicolumn{8}{|c|}{Knowledge Uncertainty-Cal-LGBM+PriorNet} \\ \hline
B & 75718 & 55346 & 20372 & 75718 & 1424 & 178 & 1246 & 1424 \\ \hline
M & 56294 & 49823 & 6471 & 56294 & 999 & 43 & 956 & 999 \\ \hline
T & 132012 & 105169 & 26843 & 132012 & 2423 & 221 & 2202 & 2423 \\ \hline
Model &\multicolumn{8}{|c|}{Expected Entropy-Cal-LGBM+NN Ensemble} \\ \hline
B & 76188 & 45771 & 30417 & 76188 & 954 & 42 & 912 & 954 \\ \hline
M & 56971 & 42770 & 14201 & 56971 & 322 & 11 & 311 & 322 \\ \hline
T & 133159 & 88541 & 44618 & 133159 & 1276 & 53 & 1223 & 1276 \\ \hline
Model &\multicolumn{8}{|c|}{Entropy of Expected-Cal-LGBM+NN Ensemble} \\ \hline
B & 76188 & 45103 & 31085 & 76188 & 954 & 38 & 916 & 954 \\ \hline
M & 56971 & 42475 & 14496 & 56971 & 322 & 9 & 313 & 322 \\ \hline
T & 133159 & 87578 & 45581 & 133159 & 1276 & 47 & 1229 & 1276 \\ \hline
Model &\multicolumn{8}{|c|}{Knowledge Uncertainty-Cal-LGBM+NN Ensemble} \\ \hline
B & 76188 & 54844 & 21344 & 76188 & 954 & 66 & 888 & 954 \\ \hline
M & 56971 & 44578 & 12393 & 56971 & 322 & 18 & 304 & 322 \\ \hline
T & 133159 & 99422 & 33737 & 133159 & 1276 & 84 & 1192 & 1276 \\ \hline
Model &\multicolumn{8}{|c|}{Cal-LGBM+NN+ICE+NCM:Probability (State-of-the-art)} \\ \hline
B & 76160 & 58118 & 18042 & 76160 & 982 & 8 & 974 & 982 \\ \hline
M & 57015 & 40630 & 16385 & 57015 & 278 & 14 & 264 & 278 \\ \hline
T & 133175 & 98748 & 34427 & 133175 & 1260 & 22 & 1238 & 1260 \\ \hline
Model &\multicolumn{8}{|c|}{Cal-LGBM+PriorNet+ICE+NCM:Probability (State-of-the-art)} \\ \hline
B & 75718 & 65951 & 9767 & 75718 & 1424 & 28 & 1396 & 1424 \\ \hline
M & 56294 & 47469 & 8825 & 56294 & 999 & 18 & 981 & 999 \\ \hline
T & 132012 & \textbf{113420} & 18592 & 132012 & 2423 & 46 & 2377 & 2423 \\ \hline
Model &\multicolumn{8}{|c|}{Cal-LGBM+NN-Ensemble+ICE+NCM:Probability (State-of-the-art)} \\ \hline
B & 76188 & 59067 & 17121 & 76188 & 954 & 7 & 947 & 954 \\ \hline
M & 56971 & 41737 & 15234 & 56971 & 322 & 13 & 309 & 322 \\ \hline
T & 133159 & 100804 & 32355 & 133159 & 1276 & \textbf{20} & 1256 & 1276 \\ \hline
Model &\multicolumn{8}{|c|}{Cal-LGBM+NN-Ensemble+ICE+NCM:Expected Entropy (our Work)} \\ \hline
B & 76188 & 49002 & 27186 & 76188 & 954 & 19 & 935 & 954 \\ \hline
M & 56971 & 38973 & 17998 & 56971 & 322 & 16 & 306 & 322 \\ \hline
T & 133159 & 87975 & 45184 & 133159 & 1276 & 35 & 1241 & 1276 \\ \hline
Model &\multicolumn{8}{|c|}{Cal-LGBM+NN-Ensemble+ICE+NCM:Entropy of Expected(Our Work)} \\ \hline
B & 76188 & 49657 & 26531 & 76188 & 954 & 21 & 933 & 954 \\ \hline
M & 56971 & 40051 & 16920 & 56971 & 322 & 15 & 307 & 322 \\ \hline
T & 133159 & 89708 & 43451 & 133159 & 1276 & 36 & 1240 & 1276 \\ \hline
Model &\multicolumn{8}{|c|}{Cal-LGBM+NN-Ensemble+ICE+NCM:Knowledge Uncertainty(our Work)} \\ \hline
B & 76188 & 49350 & 26838 & 76188 & 954 & 21 & 933 & 954 \\ \hline
M & 56971 & 39088 & 17883 & 56971 & 322 & 15 & 307 & 322 \\ \hline
T & 133159 & 88438 & 44721 & 133159 & 1276 & 36 & 1240 & 1276 \\ \hline

\end{tabulary}
\begin{flushleft}
Legend: Inst - Number of instances; CA - Correctly Accepted; CR - Correctly Rejected; IR - Incorrectly Rejected; IA - Incorrectly Accepted; B - Benign; M - Malware; T - Total number of Instances predicted correctly or incorrectly
\end{flushleft}
\end{table}

\begin{table}[htbp]
\centering
\scriptsize
\caption{Performance of UCSB tested on EMBER data set}
\label{tab:ucsb_test_EMBER_accept_reject}
\begin{tabulary}{\linewidth}{|L|L|L|L|L|L|L|L|L|}
\hline
\textbf{Data} & \multicolumn{8}{c|}{\textbf{Probability Calibration Models}} \\ \hline
\textbf{Type} & \multicolumn{4}{c|}{\textbf{Correctly Predicted}} & \multicolumn{4}{c|}{\textbf{Incorrectly Predicted}} \\ \hline
 & \textbf{Inst} & \textbf{CA} & \textbf{IR} & \textbf{T} & \textbf{Inst.} & \textbf{IA} & \textbf{CR} & \textbf{T} \\ \hline

&\multicolumn{8}{|c|}{Probability-Cal-LGBM+NN} \\ \hline
B & 15780 & 8326 & 7454 & 15780 & 86067 & 49478 & 36589 & 86067 \\ \hline
M & 207522 & 163449 & 44073 & 207522 & 2345 & 31 & 2314 & 2345 \\ \hline
T & 223302 & \textbf{171775} & 51527 & 223302 & 88412 & 49509 & 38903 & 88412 \\ \hline

&\multicolumn{8}{|c|}{Probability-Cal-LGBM+PriorNet} \\ \hline
B & 20977 & 8919 & 12058 & 20977 & 80870 & 47934 & 32936 & 80870 \\ \hline
M & 199392 & 173119 & 26273 & 199392 & 10475 & 52 & 10423 & 10475 \\ \hline
T & 220369 & 182038 & 38331 & 220369 & 91345 & 47986 & 43359 & 91345 \\ \hline

&\multicolumn{8}{|c|}{Probability-Cal-LGBM+Ensemble} \\ \hline
B & 15966 & 7722 & 8244 & 15966 & 85881 & 45679 & 40202 & 85881 \\ \hline
M & 207867 & 157607 & 50260 & 207867 & 2000 & 23 & 1977 & 2000 \\ \hline
T & 223833 & 165329 & 58504 & 223833 & 87881 & 45702 & 42179 & 87881 \\ \hline

&\multicolumn{8}{|c|}{Expected Entropy-Cal-LGBM+PriorNet} \\ \hline
B & 20977 & 7392 & 13585 & 20977 & 80870 & 40759 & 40111 & 80870 \\ \hline
M & 199392 & 131323 & 68069 & 199392 & 10475 & 577 & 9898 & 10475 \\ \hline
T & 220369 & 138715 & 81654 & 220369 & 91345 & 41336 & 50009 & 91345 \\ \hline

&\multicolumn{8}{|c|}{Entropy of Expected-Cal-LGBM+PriorNet} \\ \hline
B & 20977 & 7983 & 12994 & 20977 & 80870 & 44376 & 36494 & 80870 \\ \hline
M & 199392 & 138349 & 61043 & 199392 & 10475 & 641 & 9834 & 10475 \\ \hline
T & 220369 & 146332 & 74037 & 220369 & 91345 & 45017 & 46328 & 91345 \\ \hline

&\multicolumn{8}{|c|}{Knowledge Uncertainty-Cal-LGBM+PriorNet} \\ \hline
B & 20977 & 8137 & 12840 & 20977 & 80870 & 46196 & 34674 & 80870 \\ \hline
M & 199392 & 142908 & 56484 & 199392 & 10475 & 723 & 9752 & 10475 \\ \hline
T & 220369 & 151045 & 69324 & 220369 & 91345 & 46919 & 44426 & 91345 \\ \hline

&\multicolumn{8}{|c|}{Expected Entropy-Cal-LGBM+NN Ensemble} \\ \hline
B & 15966 & 4483 & 11483 & 15966 & 85881 & 21867 & 64014 & 85881 \\ \hline
M & 207867 & 90691 & 117176 & 207867 & 2000 & 106 & 1894 & 2000 \\ \hline
T & 223833 & 95174 & 128659 & 223833 & 87881 & 21973 & 65908 & 87881 \\ \hline

&\multicolumn{8}{|c|}{Entropy of Expected-Cal-LGBM+NN Ensemble} \\ \hline
B & 15966 & 4429 & 11537 & 15966 & 85881 & 21195 & 64686 & 85881 \\ \hline
M & 207867 & 87739 & 120128 & 207867 & 2000 & 89 & 1911 & 2000 \\ \hline
T & 223833 & 92168 & 131665 & 223833 & 87881 & 21284 & 66597 & 87881 \\ \hline

&\multicolumn{8}{|c|}{Knowledge Uncertainty-Cal-LGBM+NN Ensemble} \\ \hline
B & 15966 & 6090 & 9876 & 15966 & 85881 & 31391 & 54490 & 85881 \\ \hline
M & 207867 & 117970 & 89897 & 207867 & 2000 & 159 & 1841 & 2000 \\ \hline
T & 223833 & 124060 & 99773 & 223833 & 87881 & 31550 & 56331 & 87881 \\ \hline

&\multicolumn{8}{|c|}{Cal-LGBM+NN+ICE+NCM:Probability (State-of-the-art)} \\ \hline
B & 15780 & 8586 & 7194 & 15780 & 86067 & 20059 & 66008 & 86067 \\ \hline
M & 207522 & 102667 & 104855 & 207522 & 2345 & 38 & 2307 & 2345 \\ \hline
T & 223302 & 111253 & 112049 & 223302 & 88412 & 20097 & 68315 & 88412 \\ \hline

&\multicolumn{8}{|c|}{Cal-LGBM+PriorNet+ICE+NCM:Probability (State-of-the-art)} \\ \hline
B & 20977 & 8919 & 12058 & 20977 & 80870 & 37524 & 43346 & 80870 \\ \hline
M & 199392 & 152779 & 46613 & 199392 & 10475 & 52 & 10423 & 10475 \\ \hline
T & 220369 & 161698 & 58671 & 220369 & 91345 & 37576 & 53769 & 91345 \\ \hline

&\multicolumn{8}{|c|}{Cal-LGBM+NN-Ensemble+ICE+NCM:Probability (State-of-the-art)} \\ \hline
B & 15966 & 8847 & 7119 & 15966 & 85881 & 19998 & 65883 & 85881 \\ \hline
M & 207867 & 102964 & 104903 & 207867 & 2000 & 39 & 1961 & 2000 \\ \hline
T & 223833 & 111811 & 112022 & 223833 & 87881 & 20037 & 67844 & 87881 \\ \hline

&\multicolumn{8}{|c|}{Cal-LGBM+NN-Ensemble+ICE+NCM:Expected Entropy (Our Work) } \\ \hline
B & 15966 & 4924 & 11042 & 15966 & 85881 & 14015 & 71866 & 85881 \\ \hline
M & 207867 & 71902 & 135965 & 207867 & 2000 & 142 & 1858 & 2000 \\ \hline
T & 223833 & 76826 & 147007 & 223833 & 87881 & \textbf{14157} & 73724 & 87881 \\ \hline

&\multicolumn{8}{|c|}{Cal-LGBM+NN-Ensemble+ICE+NCM:Entropy of Expected (Our Work)} \\ \hline
B & 15966 & 5188 & 10778 & 15966 & 85881 & 15226 & 70655 & 85881 \\ \hline
M & 207867 & 74783 & 133084 & 207867 & 2000 & 138 & 1862 & 2000 \\ \hline
T & 223833 & 79971 & 143862 & 223833 & 87881 & 15364 & 72517 & 87881 \\ \hline

&\multicolumn{8}{|c|}{Cal-LGBM+NN-Ensemble+ICE+NCM:Knowledge Uncertainty (Our Work)} \\ \hline
B & 15966 & 5161 & 10805 & 15966 & 85881 & 14889 & 70992 & 85881 \\ \hline
M & 207867 & 72980 & 134887 & 207867 & 2000 & 130 & 1870 & 2000 \\ \hline
T & 223833 & 78141 & 145692 & 223833 & 87881 & 15019 & 72862 & 87881 \\ \hline

\end{tabulary}
\begin{flushleft}
Legend: Inst - Number of instances; CA - Correctly Accepted; CR - Correctly Rejected; IR - Incorrectly Rejected; IA - Incorrectly Accepted; B - Benign; M - Malware; T - Total number of Instances predicted correctly or incorrectly
\end{flushleft}

\end{table}

\subsection{Answering RQ2}

Based on the results displayed in Table \ref{classifier_performance}, the models were trained on the training sections of the datasets, with thresholds set using the validation sets and evaluations carried out on the test sets from the same datasets. The performance metrics, including accuracy, precision, and F1 score, were all relatively high, indicating that there was no significant shift in the dataset. Furthermore, when models trained on the EMBER dataset were applied to the BODMAS dataset, using thresholds determined from the EMBER validation set, the results in terms of accuracy, precision, and F1 score remained high. This consistency suggests that the BODMAS dataset did not experience a dataset shift despite being newer relative to EMBER. In response to our second research question (RQ2) about the efficacy of various predictive methods in minimizing incorrect predictions and maximizing acceptance of correct predictions in the absence of dataset shift, we analyzed the performance of each model.

In examining the Acceptance and Rejection of Predictions using  Probability thresholds, detailed in Table \ref{tab:probability_accept_reject}, it is observed that the Calibrated LGBM with NN ensemble results in the lowest acceptance of incorrect predictions in the EMBER dataset, accepting only 96 out of 1496 incorrect predictions. For the UCSB dataset, Calibrated LGBM with PriorNet shows superior performance by accepting only 84 out of 1345 incorrectly predicted instances, closely followed by the Calibrated LGBM with NN Ensemble accepting only 88 incorrect predictions. This result demonstrates the consistent efficacy of the Calibrated LGBM with NN Ensemble across both datasets when employing a Probability threshold.

From the analysis of Expected Entropy, Entropy of Expected, and Knowledge Uncertainty estimates calculated using Ensemble of Neural Network, presented in Table \ref{tab:ensemble_accept_reject} the application of Calibrated Probability and Entropy of Expected with LGBM+NN-Ensemble classifier leads to the lowest acceptance of incorrect predictions—233 out of 1496 in the EMBER dataset and 256 out of 1263 in the UCSB dataset. Comparatively, the number of incorrectly accepted predictions is higher than the probability threshold.
In comparison to the uncertainty estimate from NN Ensemble while using the uncertainty estimate from Prior Net with LGBM and PriorNet as a classifier, detailed in Table \ref{tab:prior_net_accept_reject}, the Uncalibrated Entropy of Expected method accepts more incorrect predictions—443 out of 2021 in the EMBER dataset. Conversely, in the UCSB dataset, this method accepts 352 out of 1351 incorrect predictions. It can be concluded that the uncertainty estimate from Prior Net is comparatively worse than the uncertainty estimate from NN Ensemble since the IA is relatively higher for all the uncertainty estimates.

In Inductive Conformal Evaluation, we evaluated using two different Non-Conformity Measures: a) using Probability as NCM and b) using uncertainty estimate from NN-Ensemble as NCM. Using ICE with Probability as NCM comparatively accepts the least amount of IA across different techniques like probability threshold, Uncertainty Estimate form NN-Ensemble and PriorNet in the EMBER dataset as shown in Table \ref{tab:conformal_accept_reject}. In the UCSB dataset, ICE with NCM as Probability accepts more IA than only using the Probability threshold. However, it is better than the Uncertainty Estimate from NN-Ensemble and Prior Net. While using uncertainty estimate from NN Ensemble as NCM, it performs even worse as compared to using using Probability as NCM on the same dataset split. The results of using NCM on split from the same dataset are shown in Table \ref{tab:conformal_accept_reject_uncertainity}.

We tested the BODMAS collected from the year 2019-2020 dataset using a model trained on the EMBER collected during the year 2017-2018 dataset and thresholds determined from EMBER's validation set to evaluate model performance on a dataset from a different time without dataset shift. The results, shown in \ref{tab:bodmas_test_EMBER_accept_reject}, reveal that ICE with Probability as NCM accepts the least amount of incorrect prediction compared to all the other methods closely followed by ICE with Uncertainty as NCM. Although ICE with Uncertainty estimate as NCM was accepting more incorrect predictions than probability threshold and Uncertainty estimate from NN-Ensemble in the split from the same dataset, it performed better when tested on the newer dataset. Uncertainty estimates are instrumental when a model deals with data that is very different from the data it was trained on. This deviation can be in terms of features, distributions, or underlying patterns. The unfamiliarity of the BODMAS dataset likely prompted the uncertainty-based methods to act more conservatively, and this can be seen in Table \ref{tab:conformal_accept_reject_uncertainity} where it accepts both less Correctly Predicted and Incorrectly Predicted. The performance of ICE with uncertainty estimate as NCM when there is a massive drop in F1, like in the case of UCSB, where there is a vast dataset shift, is discussed in RQ3.

\textbf{\textit{Operational Recommendation 2}}: 
Given the stable performance of the Calibrated LGBM with NN Ensemble across the EMBER, UCSB, and BODMAS datasets, particularly in terms of minimizing incorrectly accepted predictions and higher F1, this model is recommended. Additionally, in environments where data characteristics do not significantly shift, using ICE with Probability as NCM is advised for reliable decision-making. The dataset shift case is discussed in RQ3.

\subsection{Answering RQ3}\label{RQ3}
To address RQ3, focusing on the impact of dataset shift on uncertainty estimation methods, we evaluated models trained on the EMBER dataset against the UCSB dataset, which consists of Packed PE files. This scenario introduces a covariate shift, \(P_{train}(X_{unpacked}) \neq P(X_{packed})\), due to the alteration in input feature distribution resulting from the packing process, as detailed in the dataset description. The reduction in F1 score to 72.56 from 99.53 shown in \ref{tab:conformal_accept_reject_uncertainity} clearly shows the performance impact due to the dataset shift.

The analysis presented in Table \ref{tab:ucsb_test_EMBER_accept_reject} reveals insights into the effectiveness of accepting and rejecting prediction methodologies applied across UCSB data tested on EMBER. Notably, the models applying the Entropy of Expected metric from the NN-Ensemble without utilizing Inductive Conformal Evaluation (ICE) resulted in 21,284 incorrect predictions out of a total of 87,881. This performance slightly surpasses that of the Expected Entropy metric from the same ensemble, which accounted for 21,973 incorrect predictions.

\begin{figure}[htbp]
    \centering
    \includegraphics[width=1.0\linewidth]{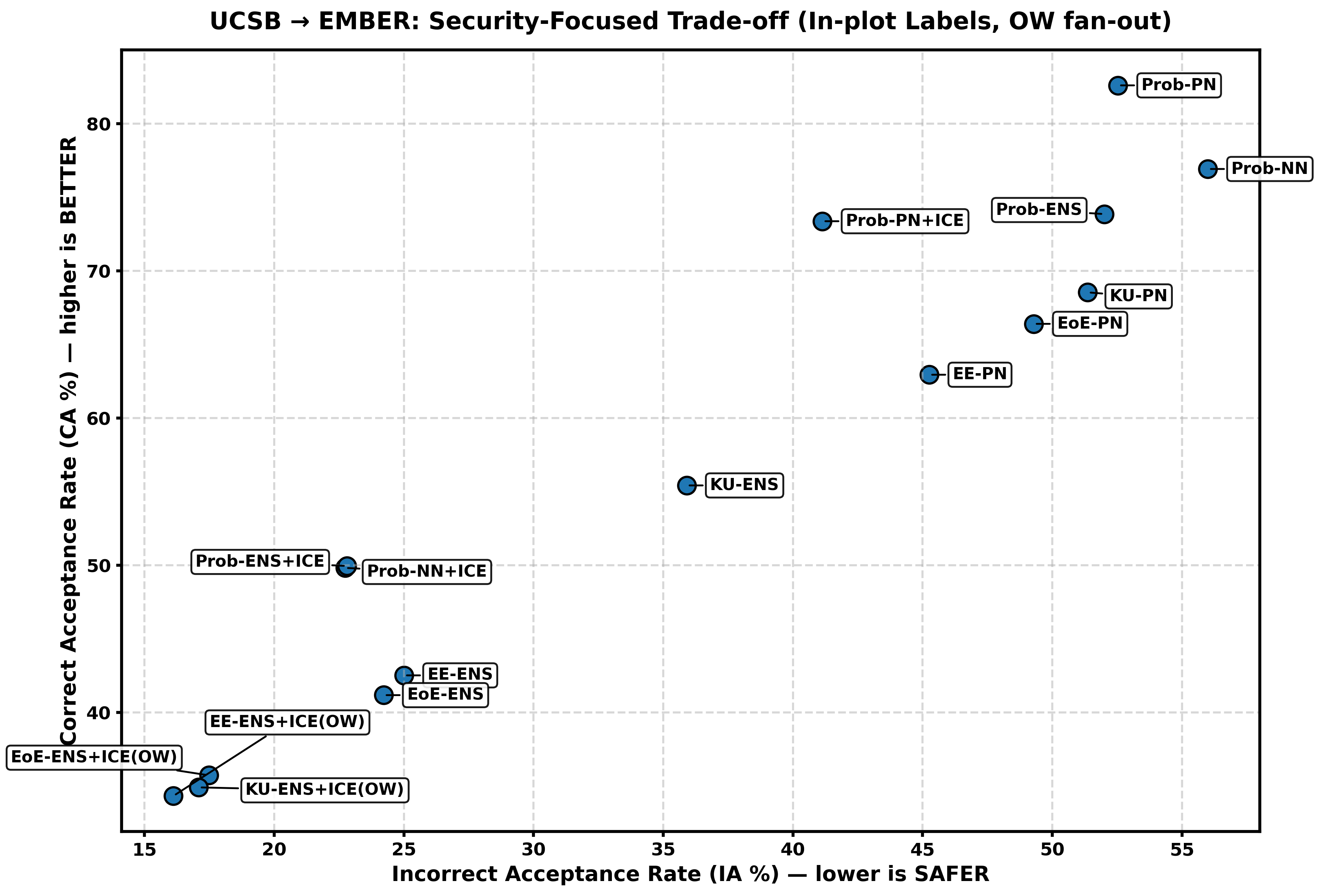}
    \caption{Performance assessment under severe out-of-distribution (OOD) shift: The UCSB dataset was tested on the EMBER dataset. The plot shows the trade-off between Correct Acceptance Rate (CA\%) and Incorrect Acceptance Rate (IA\%). 
    Lower IA\% corresponds to safer behaviour (fewer incorrect predictions accepted), while higher CA\% indicates better retention of correct predictions. 
    Labels are embedded inside the plot to identify each model. \textbf{Label mapping:} Prob-NN = Probability-Cal-LGBM+NN; Prob-PN = Probability-Cal-LGBM+PriorNet; Prob-ENS = Probability-Cal-LGBM+Ensemble; 
    EE-PN = Expected Entropy-Cal-LGBM+PriorNet; EoE-PN = Entropy of Expected-Cal-LGBM+PriorNet; KU-PN = Knowledge Uncertainty-Cal-LGBM+PriorNet; 
    EE-ENS = Expected Entropy-Cal-LGBM+NN Ensemble; EoE-ENS = Entropy of Expected-Cal-LGBM+NN Ensemble; KU-ENS = Knowledge Uncertainty-Cal-LGBM+NN Ensemble; 
    Prob-NN+ICE = Cal-LGBM+NN+ICE+NCM:Probability (SOTA); Prob-PN+ICE = Cal-LGBM+PriorNet+ICE+NCM:Probability (SOTA); Prob-ENS+ICE = Cal-LGBM+NN-Ensemble+ICE+NCM:Probability (SOTA); 
    EE-ENS+ICE(OW) = Cal-LGBM+NN-Ensemble+ICE+NCM:Expected Entropy (Our Work); 
    EoE-ENS+ICE(OW) = Cal-LGBM+NN-Ensemble+ICE+NCM:Entropy of Expected (Our Work); 
    KU-ENS+ICE(OW) = Cal-LGBM+NN-Ensemble+ICE+NCM:Knowledge Uncertainty (Our Work).}
    \label{fig:ucsb_ember_tradeoff}
\end{figure}

Further examination shows that calibrated models such as LGBM+NN and the LGBM+NN Ensemble using ICE with Probability as the Non-Conformity Measure (NCM) demonstrate comparable effectiveness. They accept 20,097 and 20,037 incorrect predictions, respectively, which is less than half of other techniques that don't use uncertainty estimates from NN Ensemble. However, our proposed method using ICE combined with uncertainty estimates as NCM significantly reduces the number of incorrect predictions compared to the one that uses probability as NCM. Specifically, using Expected Entropy, Entropy of Expected, and Knowledge Uncertainty as NCMs with ICE leads to even fewer incorrect acceptances, tallying at 14,157, 15,364, and 15,019, respectively. These results highlight the enhanced precision and adaptability of ICE when integrated with sophisticated uncertainty metrics, proving it to be more effective in managing incorrect predictions, particularly in scenarios involving dataset shifts.   The improved performance of models employing Inductive Conformal Evaluation (ICE) with advanced uncertainty metrics such as Expected Entropy, Entropy of Expected, and Knowledge Uncertainty suggests that these uncertainty metrics provide a more refined assessment of prediction confidence, particularly in environments marked by uncertainty, thereby enabling the models to reject Incorrect predictions better. To further illustrate these results, Figure \ref{fig:ucsb_ember_tradeoff} visualizes the trade-off between Correct Acceptance Rate (CA\%) and Incorrect Acceptance Rate (IA\%) for UCSB tested on EMBER. The figure clearly shows that our proposed ICE-based approaches with advanced uncertainty metrics (Expected Entropy, Entropy of Expected, and Knowledge Uncertainty) achieve substantially lower IA\% while maintaining competitive CA\%. In contrast, baseline probability-based ICE models cluster toward higher IA\%, reflecting their tendency to accept more incorrect predictions under dataset shift. This visualization complements the tabular results by highlighting the security-critical improvement of our method: fewer incorrect acceptances without sacrificing too much correct retention. 

The Correct Acceptance Rate (CA\%) and Incorrect Acceptance Rate (IA\%) are defined as:

\begin{equation}
    CA\% = \frac{CA}{\text{Correct Total}} \times 100, 
\end{equation}
\begin{equation}
    IA\% = \frac{IA}{\text{Incorrect Total}} \times 100
\end{equation}

where $CA$ denotes the number of correctly accepted predictions, $\text{Correct Total}$ is the total number of correct predictions, 
$IA$ is the number of incorrectly accepted predictions, and $\text{Incorrect Total}$ is the total number of incorrect predictions.

An intriguing aspect of the analysis is the distribution of misclassifications and acceptance of incorrect predictions, particularly concerning benign software. The data suggests that benign software, often packed or obfuscated to protect intellectual property or prevent license abuse, is more likely to be misclassified. Despite the dataset shift primarily affecting benign samples, all evaluated techniques demonstrated a robust capability to reject incorrect malware predictions or accept fewer incorrectly predicted malware instances, indicating that the models effectively handle dataset shifts, especially in distinguishing malware.

\textbf{\textit{Operational Recommendation 3}}: In scenarios where dataset shift occurs, the incorporation of advanced uncertainty metrics like Expected Entropy, Entropy of Expected, and Knowledge Uncertainty as Non-Conformity Measures (NCM) within the Inductive Conformal Evaluation (ICE) framework has demonstrated superior effectiveness in minimizing incorrect predictions, compared to the state of the art method that employs Probability as NCM. Consequently, it is recommended that uncertainty estimates from NN Ensembles within the ICE framework be utilized. This approach not only enhances the accuracy of the predictions but also ensures robustness against variations in data characteristics.

\subsection{Answering RQ4}

Our experiments have demonstrated that while the Calibrated LightGBM model integrated with a Prior Network yields the fewest incorrect predictions when utilizing probability or ICE with Probability as NCM on consistent UCSB dataset splits, its efficacy declines with any shift in the dataset. Despite its strong performance on stable UCSB splits, it only matches that of the Calibrated LGBM with an NN Ensemble under similar conditions. We observed that the Prior Network's performance heavily depends on exposure to a broad spectrum of adversarial or out-of-distribution (OOD) data, as it is specifically trained to recognize a uniform distribution for OOD data. This requirement presents a significant challenge, particularly in malware detection, where generating comprehensive OOD data can be complex.

The ICE framework with Probability as NCM excels in environments without dataset shifts but when dataset shifts occur, integrating uncertainty estimates from an NN ensemble within the ICE framework has shown to be the most effective approach.

\textbf{\textit{Operational Recommendation 4}}: Based on our findings, we conclude that relying solely on a single deterministic model, even one that is well-calibrated, is less effective than using ensemble methods for uncertainty estimation, particularly in the presence of dataset shifts. The superior performance of the ensemble model underscores its robustness and flexibility, making it more suitable for varying conditions. Therefore, it is advisable to prefer ensemble methods over single deterministic models, especially in applications where dataset variability is a concern. This approach not only enhances prediction accuracy but also ensures greater reliability across different operational scenarios.

While this study focuses on Windows PE malware using tree-based and neural classifiers, the proposed uncertainty estimation framework is not restricted to this application domain. The integration of ensemble-based uncertainty with conformal prediction is classifier-agnostic and can be applied to other domains, such as Android malware detection or even medical image classification. We intentionally selected PE malware to scope the evaluation for tractability, but extending the methodology to Android datasets (e.g., Drebin, Androzoo) is a promising direction for future work that may further validate its generalizability.

\section{Limitations and Discussion}
Our evaluation demonstrates strong performance under dataset shift (e.g., EMBER-trained models tested on UCSB and BODMAS), where ensemble-based uncertainty measures combined with Inductive Conformal Evaluation (ICE) substantially reduce incorrectly accepted predictions. Nevertheless, several limitations remain. First, under extreme shifts—where future malware families differ entirely from those represented in the training and calibration sets—the reliability of both probability- and ensemble-based uncertainty estimates may degrade. This is a known limitation of methods that assume at least partial overlap between calibration and deployment distributions.

Second, while our harmonic threshold optimization balances correctly accepted and correctly rejected predictions on the validation set, its generalization may weaken if the validation data fail to capture the characteristics of future shifts.

Finally, our study does not explicitly evaluate adversarial attacks, such as carefully crafted feature perturbations designed to evade detection. Prior work has shown that adversarially generated samples can compromise malware classifiers even when uncertainty estimation is used, highlighting the need for integrating adversarial robustness into our framework. Future work could address these challenges by incorporating adversarial training, dynamic threshold calibration, or data augmentation with synthetic attacks to further harden the system.

\section{Conclusion}\label{conclusion}
Through our empirical study, we rigorously evaluated how well different known Windows PE malware detectors accept or reject predictions. The study using three Windows PE malware datasets, EMBER, UCSB and BODMAS, revealed that calibrating probabilities significantly improves the rejection of both malware and benign predictions. Specifically, when there is no dataset shift, the calibrated probability threshold and Inductive Conformal Evaluation (ICE) with probabilities as Non-Conformity Measure (NCM) consistently perform better across datasets than other uncertainty estimation methods. However, in situations with dataset shifts, our proposed method, which uses the uncertainty estimates from the NN Ensemble as a nonconformity measure, proved more effective than the probability threshold, ICE with probability estimate as NCM, and uncertainty estimates from the PriorNet. Our method was able to reject 73,724, 72,517 and 72,8269 incorrect predictions out of 87,881 incorrect predictions using Expected Entropy, Entropy of Expected and Knowledge Uncertainty as NCM in the ICE framework along with Calibrated LBM+NN-Ensemble, respectively.

In comparison, the state-of-the-art method ICE with probability estimate using Calibrated LGBM+NN-Ensemble rejected 67,884 incorrect predictions out of 87,881 incorrect predictions \cite{barbero2022transcending}. It is important to note that although the PriorNet is considered a highly efficient alternative to the ensemble approach, it did not outperform the uncertainty estimates from the NN ensemble. This underperformance is likely due to the inherent challenges in generating suitable out-of-distribution (OOD) data, underscoring the complexities of applying uncertainty estimation in dynamic environments like malware detection.

Beyond these experimental results, our findings also carry practical implications for real-world cybersecurity operations. In practice, uncertainty-aware conformal prediction can act as a triage mechanism: high-confidence predictions can be handled automatically within security pipelines, while low-confidence or high-uncertainty predictions can be routed for manual analyst review. This selective prediction framework reduces the risk of false acceptance and helps security teams prioritize attention where it is most needed. Moreover, the approach integrates naturally into malware analysis and intrusion detection workflows, offering an additional safeguard in environments where operational costs of misclassification are high.

Our work focuses on improving decision reliability via uncertainty-aware rejection under dataset shifts. We did not incorporate frequency-based, feature-based, or adversarial defense mechanisms. Integrating these modern defense approaches alongside uncertainty estimation offers a promising direction for future research.

%{\appendices
%\section*{Proof of the First Zonklar Equation}
%Appendix one text goes here.
% You can choose not to have a title for an appendix if you want by leaving the argument blank
%\section*{Proof of the Second Zonklar Equation}
%Appendix two text goes here.}

%\section{References Section}
 
 % argument is your BibTeX string definitions and bibliography database(s)
%\bibliography{IEEEabrv,../bib/paper}

{\appendix[]}
\begin{figure}[H]
     % \begin{minipage}{0.46\textwidth}
        \centering
        \includegraphics[width=0.35\textwidth]{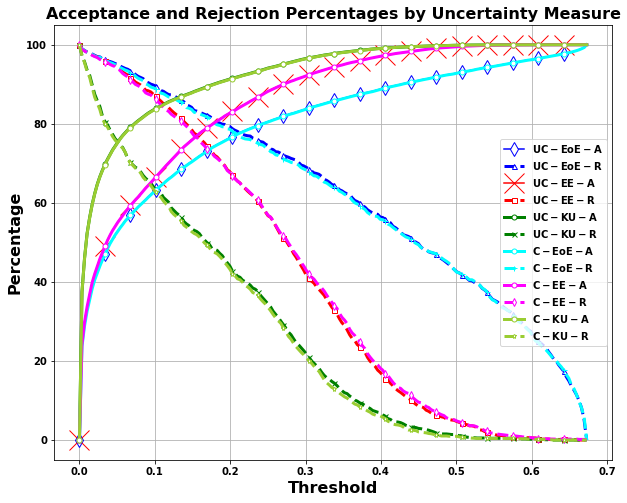}
        \caption{Correctly Accepted and Correctly Rejected Trade-off Graph using Uncertainity Estimates from NN ensemble on UCSB Test dataset. Where: UC: Uncalibrated, C-Calibrated, Correctly Accepted: A, Correctly Rejected: R, EE: Expected Entropy, EoE: Entropy of Expected and KU: Knowledge Uncertainity  }
        \label{fig:ensemble-accept-reject-UCSB}
    % \end{minipage}\hfill

\end{figure}

 \begin{figure}[H]
 % \begin{minipage}{0.48\textwidth}
    \centering
    \includegraphics[width=0.35\textwidth]{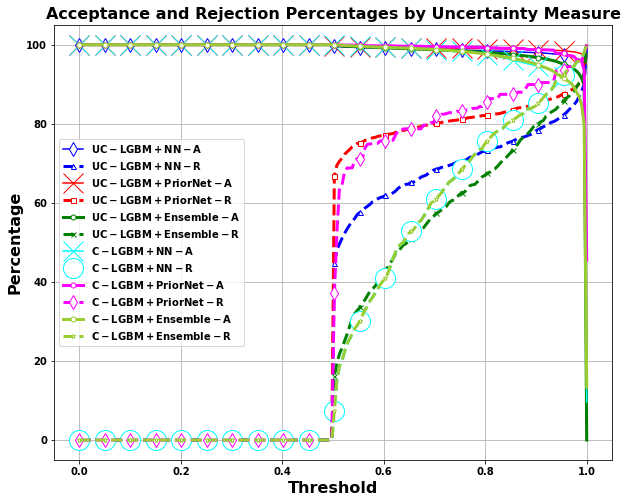}
    \caption{Correctly Accepted and Correctly Rejected Trade-off Graph using a range of Probability threshold on EMBER Test dataset. Where, UC: Uncalibrated, C-Calibrated, Correctly Accepted:A, Correctly Rejected: R}
    \label{fig:Prob-accept-reject-EMBER}
\end{figure}
%     \end{minipage}
\begin{figure}[H]
    \centering
    \includegraphics[width=0.35\textwidth]{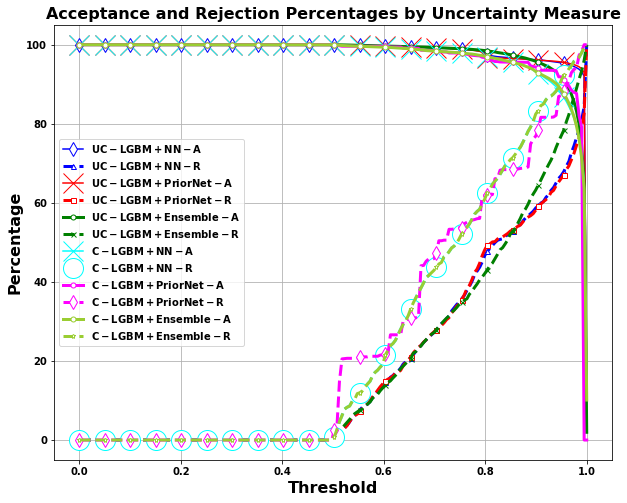}
    \caption{Correctly Accepted and Correctly Rejected Trade-off Graph using a range of Probability threshold on UCSB Test dataset. Where, UC: Uncalibrated, C-Calibrated, Correctly Accepted:A, Correctly Rejected: R}
    \label{fig:prob-accept-reject-ucsb}
\end{figure}

\begin{figure}[H]
        \centering
        \includegraphics[width=0.35\textwidth]{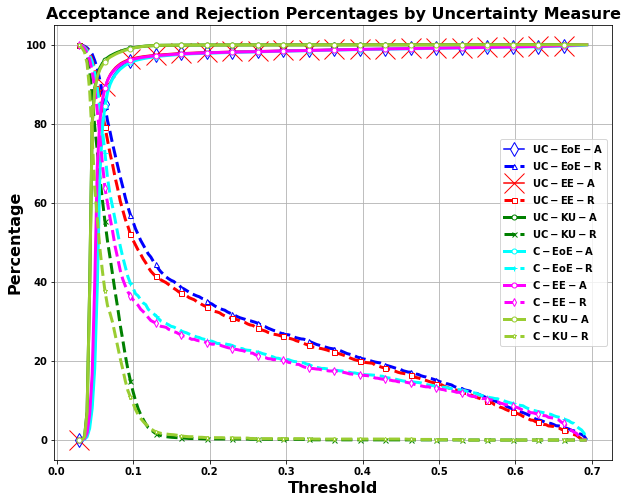}
        \caption{Correctly Accepted and Correctly Rejected Trade-off Graph using Uncertainity Estimates from Prior Net on EMBER Test dataset. Where: UC: Uncalibrated, C-Calibrated, Correctly Accepted:A, Correctly Rejected: R, EE: Expected Entropy, EoE: Entropy of Expected and KU: Knowledge Uncertainity }
        \label{fig:prior-accept-reject-EMBER}
\end{figure}
\begin{figure}[H]
        \centering
        \includegraphics[width=0.35\textwidth]{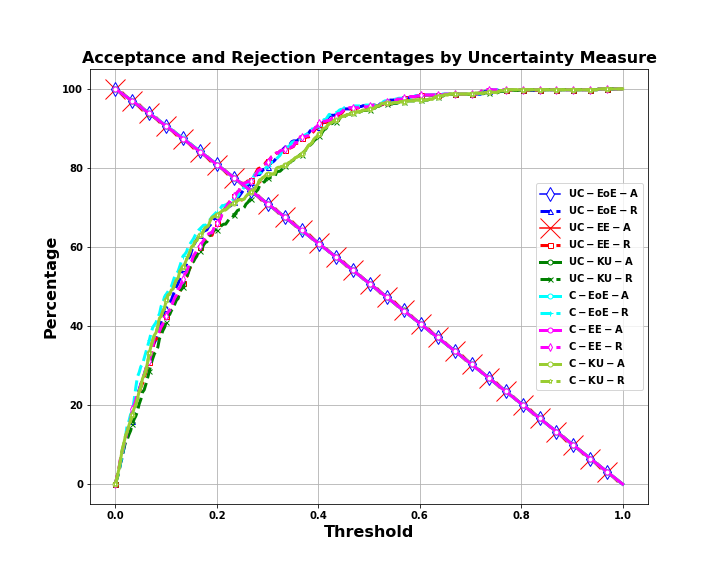}
        \caption{Correctly Accepted and Correctly Rejected Trade-off Graph for ICE with Uncertainity estimate from NN Ensemble as NCM. Where: UC: Uncalibrated, C-Calibrated, Correctly Accepted:A, Correctly Rejected: R, EE: Expected Entropy, EoE: Entropy of Expected and KU: Knowledge Uncertainity }
        \label{fig:prior-accept-reject-ucsb}
\end{figure}

\begin{figure}

        \centering
        \includegraphics[width=0.35\textwidth]{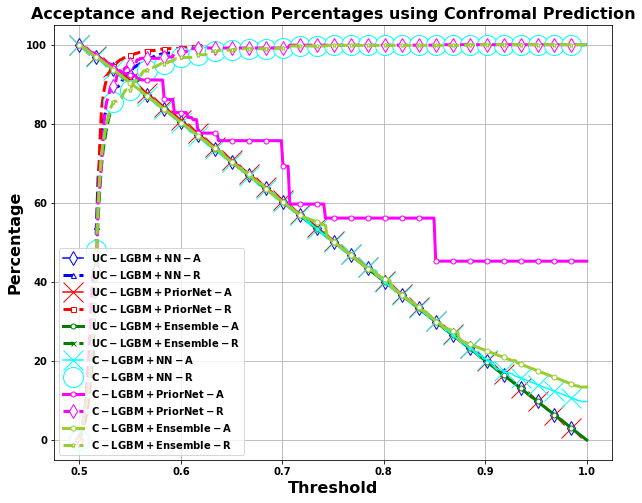}
        \caption{Correctly Accepted and Correctly Rejected Trade-off Graph using a range of threshold for ICE with probability as NCM on EMBER Test dataset. Where, UC: Uncalibrated, C: Calibrated, Correctly Accepted:A, Correctly Rejected: R}
        \label{fig:conformal-accept-reject-EMBER}
\end{figure}
\begin{figure}
        \centering
        \includegraphics[width=0.38\textwidth]{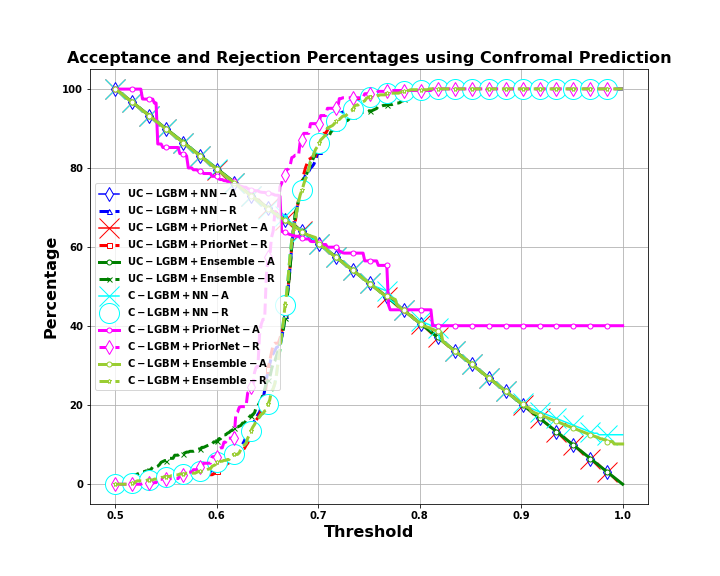}
        \caption{Correctly Accepted and Correctly Rejected Trade-off Graph using a range of threshold for ICE with probability as NCM on UCSB Test dataset. Where, UC: Uncalibrated, C: Calibrated, Correctly Accepted: A, Correctly Rejected: R}
        \label{fig:conformal-accept-reject-ucsb}
\end{figure}

%\newpage

% \newline
%\section{Biography Section}
% If you have an EPS/PDF photo (graphicx package needed), extra braces are
%  needed around the contents of the optional argument to biography to prevent
%  the LaTeX parser from getting confused when it sees the complicated
%  $\backslash${\tt{includegraphics}} command within an optional argument. (You can create
%  your own custom macro containing the $\backslash${\tt{includegraphics}} command to make things
%  simpler here.)
 
\vspace{11pt}

\vspace{-33pt}
\begin{IEEEbiography}[{\includegraphics[width=1in,height=1.25in,clip,keepaspectratio]{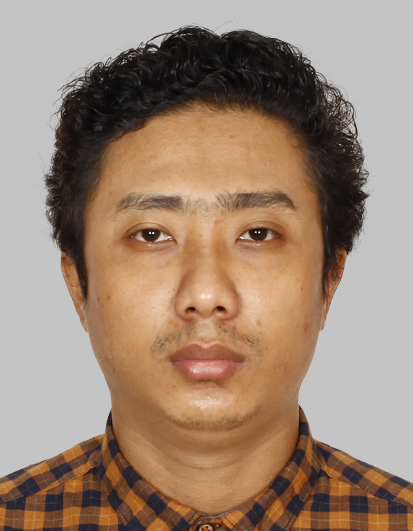}}]{Rahul Yumlembam} has completed the PhD degree in Computer and Information Sciences at Northumbria University, UK. He finished BTech degree in Computer Science and Engineering and MTech degree in Computer Science and Engineering (Artificial Intelligence) in India. He was a Project Fellow at IIT, Guwahati, India. His research interest includes Machine Learning, Deep learning, Cyber Security using AI, Big Data, Brain-Computer Interface, Text Mining, and Image Processing.
% Use $\backslash${\tt{begin\{IEEEbiography\}}} and then for the 1st argument use $\backslash${\tt{includegraphics}} to declare and link the author photo.
% Use the author name as the 3rd argument followed by the biography text.

\end{IEEEbiography}
% \vskip -3\baselineskip plus -1fil
\begin{IEEEbiography} [{\includegraphics[width=1in,height=1.25in,clip,keepaspectratio]{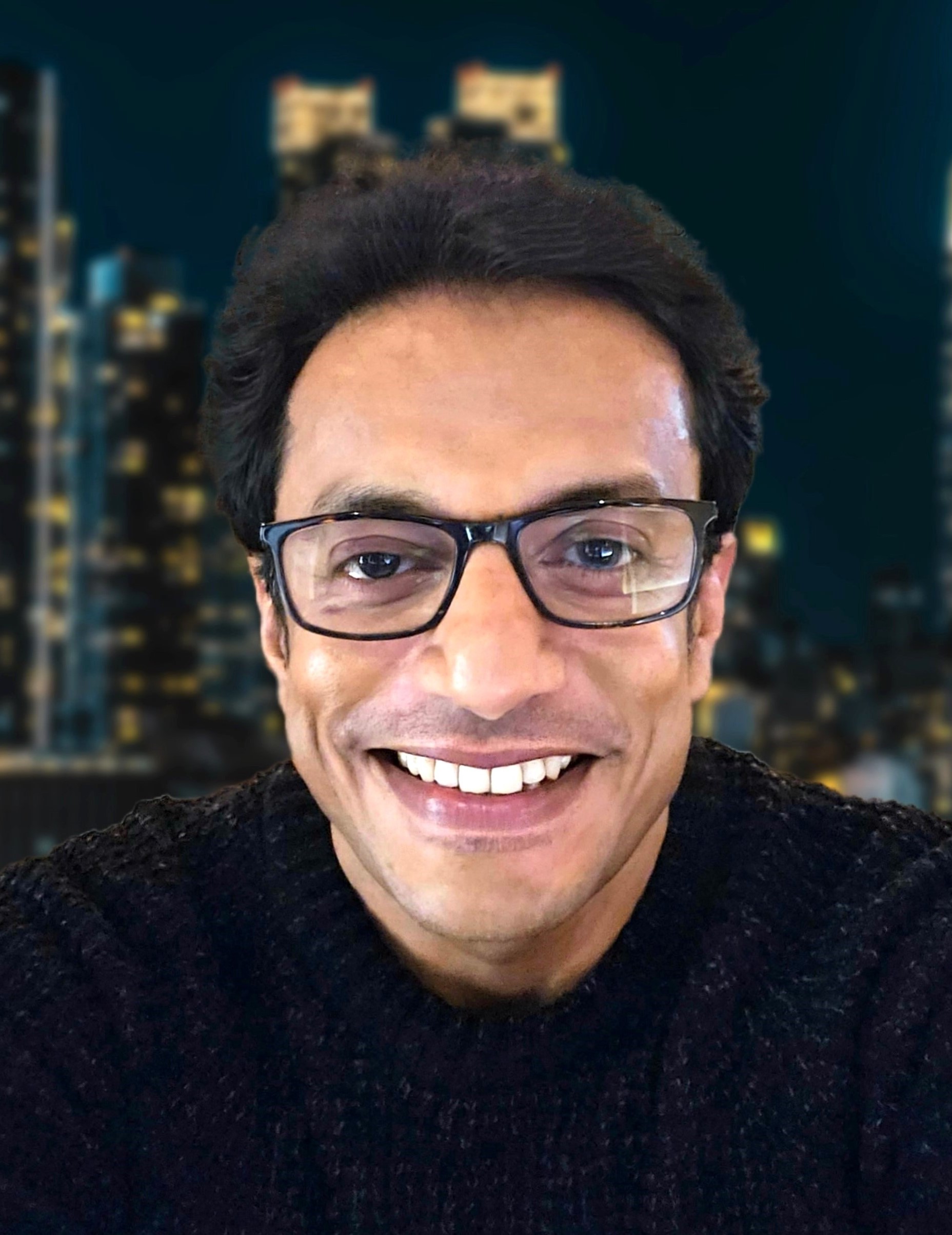}}]{Biju Issac} received the BE degree in Electronics and Communications Engineering, the Master of Computer Applications (MCA) degree, and the PhD degree in Networking and Mobile Communications. He is an Associate Professor in Northumbria University, UK. He is research active and has authored more than 100 refereed conference papers, journal articles, and book chapters. His research interests include Networks, Cybersecurity, Applied Machine Learning, Hate Speech/Fake news detection etc.
% Use $\backslash${\tt{begin\{IEEEbiography\}}} and then for the 1st argument use $\backslash${\tt{includegraphics}} to declare and link the author photo.
% Use the author name as the 3rd argument followed by the biography text.
\end{IEEEbiography}
% \vskip -3\baselineskip plus -1fil
\begin{IEEEbiography}[{\includegraphics[width=1in,height=1.25in,clip,keepaspectratio]{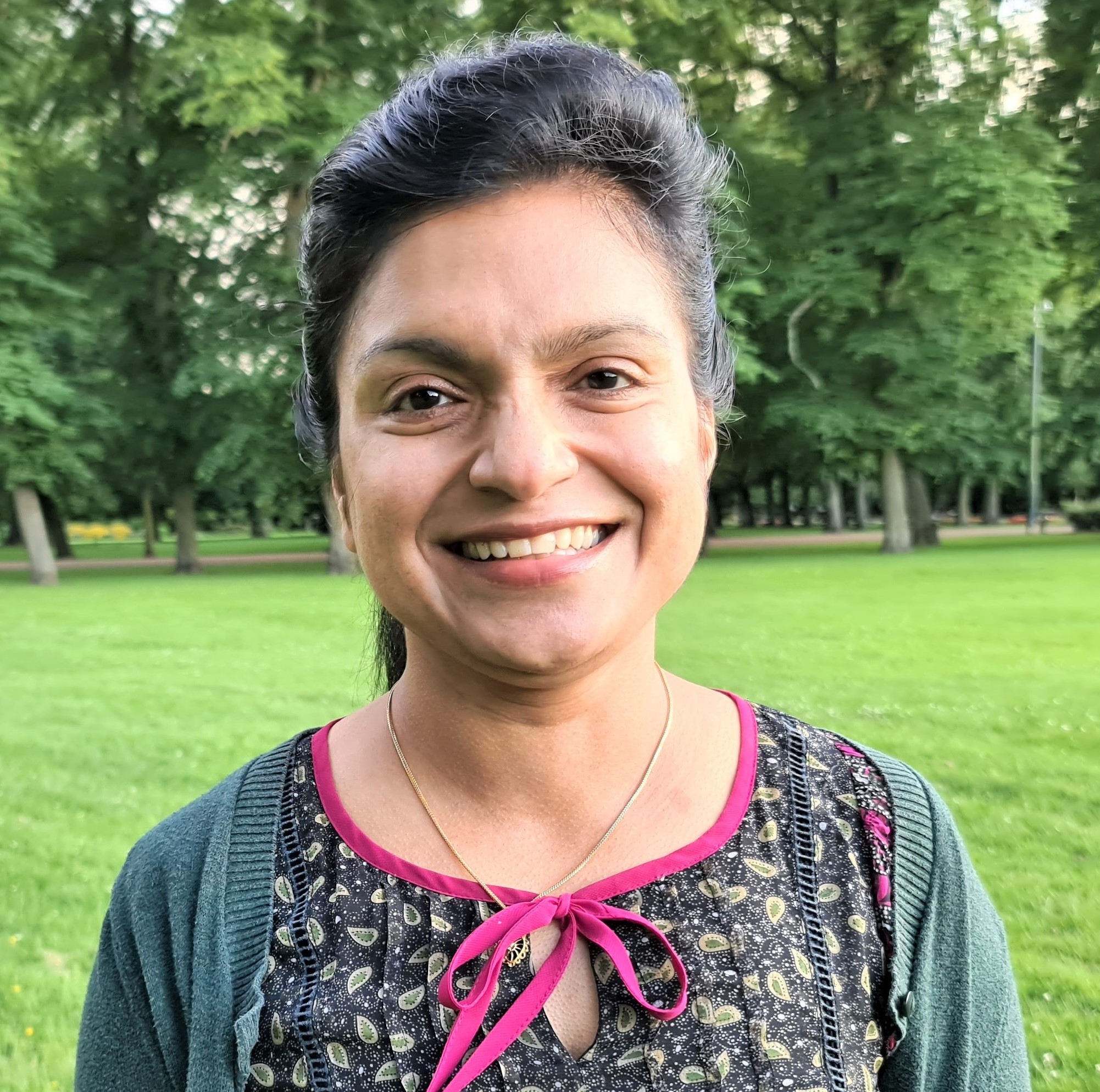}}]{Seibu Mary Jacob} received the BSc and MSc degrees in Mathematics, the Post Graduate Diploma in Computer Applications (PGDCA) degree, the Bachelor’s degree in Mathematics Education (BEd), and the PhD degree in Mathematics Education. She is currently a Senior Lecturer teaching Engineering Mathematics in Teesside University, UK. She has authored more than 30 research publications as book chapters, journal articles, and conference papers. 
% Use $\backslash${\tt{begin\{IEEEbiography\}}} and then for the 1st argument use $\backslash${\tt{includegraphics}} to declare and link the author photo.
% Use the author name as the 3rd argument followed by the biography text.
\end{IEEEbiography}
% \vskip -3\baselineskip plus -1fil

\vspace{11pt}

% \bf{If you will not include a photo:}\vspace{-33pt}
% \begin{IEEEbiographynophoto}{John Doe}
% Use $\backslash${\tt{begin\{IEEEbiographynophoto\}}} and the author name as the argument followed by the biography text.
% \end{IEEEbiographynophoto}

\vfill

\end{document}